\documentclass[aps,pre,onecolumn,showpacs,showkeys,a4paper]{revtex4}
\input epsf  
\usepackage{graphicx} 
\usepackage{amsmath}
\usepackage{amssymb}
\usepackage{amscd}

\begin{document} 

\title{Anharmonic oscillator driven by  additive Ornstein-Uhlenbeck noise}
\author{Kirone Mallick}
  \affiliation{Service de Physique Th\'eorique, Centre d'\'Etudes de Saclay,
 91191 Gif-sur-Yvette Cedex, France}
  \email{mallick@spht.saclay.cea.fr}
 \author{Philippe Marcq}
  \affiliation{Institut de Recherche sur les Ph\'enom\`enes Hors \'Equilibre,
  Universit\'e de Provence,
  49 rue Joliot-Curie, BP 146, 13384 Marseille Cedex 13, France}
  \email{marcq@irphe.univ-mrs.fr}
  \date{October 22, 2004}

\begin{abstract}
  We present an analytical study of  a nonlinear oscillator 
 subject to   an  additive  Ornstein-Uhlenbeck noise.
 Known results are mainly  perturbative  and  are  restricted 
 to  the large  dissipation limit (obtained by neglecting  the inertial term)  
 or to a   quasi-white noise ({\it i.e.},  a noise with  vanishingly
  small correlation   time).  
 Here,  in contrast, we study the small dissipation case  
 (we retain  the inertial term) and  consider a noise with finite correlation time.
 Our  analysis is  non perturbative  and 
  based on   a recursive  adiabatic elimination scheme: 
  a  reduced  effective Langevin dynamics for the slow  action variable is  obtained
  after averaging out the fast angular variable.   
 In the conservative case,  we show that the physical observables
 grow algebraically with time and  calculate 
 the associated anomalous scaling 
 exponents and  generalized diffusion constants.  In the case
 of  small dissipation, we  derive an analytic 
 expression  of  the stationary  Probability Distribution Function (P.D.F.) 
 which differs from  the canonical
 Boltzmann-Gibbs distribution.  Our results  are in excellent agreement with  
  numerical simulations.

 \end{abstract}
\maketitle

 \section{Introduction}

   A theoretical description of  the intrinsic  thermal fluctuations
 induced by the coupling of a given dynamical system  with  
 a heat  bath can be obtained by adding 
  a noise term to the  equations of motion.   
   From   this point of view   that  originates
 from Langevin's treatment of Brownian motion,  the
 effect of the  heat  bath is modeled as a  random force 
 that transforms  a  deterministic  motion into a stochastic
 dynamics in  phase space. 
 The physics of the system is then described by a 
  probability distribution function  (P.D.F.) 
 and the information about  the observables is 
 of statistical nature. When the additive noise is white 
 ({\it  i.e.}, the  correlation time of the noise 
 is smaller than any intrinsic
 time scale of the system) and its amplitude satisfies 
 the  fluctuation-dissipation relation, the stationary P.D.F.
 is identical to  the canonical Boltzmann-Gibbs
  distribution \cite{vankampen,gardiner}.

 A paradigm of  a 
  stochastic dynamical system  for a   quantitative study of 
 the interplay between randomness and  nonlinearity is provided 
 by  a  particle trapped in a nonlinear
 confining  potential  and  subject to a  random   noise 
  \cite{strato,landaMc}.  We showed in 
  \cite{philkir1,philkir2,philkir3} 
 that in the limit of a  vanishingly small
  damping rate, such a  particle exhibits   
 anomalous diffusion   with exponents that  depend  on the
  form  of the confining potential at infinity.
 For a  non-zero damping rate,  this  anomalous diffusion
   occurs  as  intermediate time asymptotics:
  the particle diffuses in phase space
 and absorbs energy from the noise until the dissipative time
 scale is reached and  the physical observables become
 stationary.    For  an additive Gaussian  white noise,
 it  was  shown  explicitly  \cite{philkir2}  that,  for times
  larger than the
 inverse damping rate,  the intermediate time   P.D.F.
  matches the canonical distribution.
  For   colored noise,  we  observed from numerical simulations  
   that the anomalous growth  exponents 
 are  different from those calculated for  white noise. 
 This behavior strongly suggests that 
  the long   time asymptotics   in the case 
 of colored noise is not  identical to
 the canonical Boltzmann-Gibbs distribution.
 The physical reason for this change of behavior   is the following:
     the  intrinsic  period   of a  nonlinear oscillator   is a 
  decreasing  function   of its  energy and when  the amplitude
 of the oscillations grows  the  period    eventually  becomes  shorter than 
  the  correlation time of the noise. In that case,  
  destructive interference  between the fast  variable and the noise 
   suppresses  the energy transfer from  the noise
   to the system and the diffusion slows down. In this regime, 
  the correlation time of the noise ceases to be the shortest
  time scale in the system and the noise can not be treated
  perturbatively as white. Therefore, usual perturbative
  calculations based on small correlation
 time expansions  \cite{sancho1,sancho2,weinstein}  cannot  predict
 the correct colored noise scalings (this was shown explicitely 
 for the pendulum with multiplicative noise in  \cite{philkir4}).
  Besides,  
   the  averaging  procedure  used  for the
   white noise fails for colored noise  and in  \cite{philkir2}
   we could only infer 
  the colored noise  exponents using   qualitative scaling arguments.

 Our aim, in this work, is to perform an  analytical study of  
 nonlinear oscillator subject to 
 an additive  Ornstein-Uhlenbeck noise and to derive
 quantitative results for its  long time behavior. 
 We need to generalize the averaging method used for white noise
 in  \cite{philkir1,philkir2}  so that it can be applied   to
 a regime where  the correlation time of the noise is not
 the shortest time scale  in  the system. 
 We define, in a recursive manner,
 new coordinates on phase space that embody, order by order,
 correlations between the noise and the  fast  angular  variable
  and then   average out   the fast  variable. 
 This technique allows us to obtain  an analytical expression 
 for the long  time asymptotic behavior of the 
 P.D.F. of the system, to  derive  the  anomalous scaling 
 exponents and  to  calculate generalized diffusion constants. 
 In particular, it is shown explicitly 
 that the long time  P.D.F.  of  a nonlinear oscillator 
subject to additive colored noise  is not the canonical
 Boltzmann-Gibbs distribution. 
 Our results are  systematically  compared  with numerical simulations. 

 The outline of this paper is as follows.  In section \ref{sec:review},
 we review  results  derived in  \cite{philkir2}  and which  are required for
 the present work. In section \ref{sec:averaging} 
 we first  explain why the straightforward averaging technique
 fails for colored noise,    provide a  heuristic calculation
   that explains the idea underlying the correct
  averaging scheme  and  develop it analytically. An effective
 averaged Langevin  system for the slow variables is derived  in the 
 subsection   \ref{sec:averagedLang}. In section \ref{sec:results}, we derive 
 analytical expressions  for the P.D.F., the anomalous
 diffusion exponents and the generalized diffusion constants;
 we also extend  our calculations to a weakly dissipative system
  and compare our results with numerical simulations.
   Section \ref{sec:conclusion} is devoted to  concluding remarks.
  Technical mathematical details are given in the appendices. 
 In particular, in appendix \ref{sec:appBFPE}, we discuss the  
 `Best Fokker-Planck Equation' (B.F.P.E.)
 approximation \cite{lindcol1,lindcol2}, based on 
  partial resummation   to all orders in the  correlation time $\tau$     
  of a perturbative  expansion  for  small noise amplitude.

\section{Review of earlier  results}
\label{sec:review}
   Consider a nonlinear oscillator of amplitude $x(t)$, trapped
 in  a  confining potential ${\mathcal U}(x)$  and subject to an additive 
 noise $\xi(t)$~:
 \begin{equation}
   \frac{\textrm{d}^2 }{\textrm{d} t^2}x(t) 
   = - \frac { \partial{\mathcal U}(x)}{\partial x}  +  \xi(t) \,.
 \label{dynamique}
\end{equation}
 The statistical properties of the random function $\xi(t)$
 need not be specified at this stage. 
 We restrict our analysis to the case where the potential
  ${\mathcal U}$ is  
  an even polynomial in $x$  for  $ |x| \rightarrow \infty$. 
   A suitable rescaling of $x$ allows us to write
 \begin{equation}
     {\mathcal U}   \sim \frac{ x^{2n}}{2n}
  \,\, \hbox{ with } \,\, n \ge  2 \,. 
 \label{infU}
\end{equation}
As the amplitude $x(t)$ of the oscillator grows with time,
 only the asymptotic behavior of ${\mathcal U}(x)$ for 
$ |x| \rightarrow \infty$ is  relevant and Eq.(\ref{dynamique})
 reduces to
  \begin{equation}
   \frac{\textrm{d}^2 }{\textrm{d} t^2}x(t)  + x(t)^{2n-1}  =  \xi(t)  \,.
 \label{dyn2}
\end{equation}
We emphasize that all stochastic differential equations involving 
white noise are interpreted in the Stratonovich sense, that allows 
the use of ordinary differential calculus for the change of variables. 
We review   here some  previously derived   results that are required  
for  the   present work.

 \subsection{Energy-angle coordinates}
\label{sec:white:aa}

 A key feature of Eq.(\ref{dyn2}) is that 
  its  underlying deterministic system
  (obtained by setting $\xi \equiv  0$), is {\it integrable}.
 The associated  energy and the angle variable  are given by 
 \cite{philkir1,philkir2}
\begin{equation}
  E =  \frac{1}{2}\dot x^2 + \frac{1}{2n} x^{2n} \, \,\, ,
  \;\;\; \hbox{ and } \;\;\;   \phi  = 
 \frac{ \sqrt{n}} { (2n)^{1/2n} } \int_0^ { {x}/{E^{{1}/{2n}}} }
   \frac{{\textrm d}u}{\sqrt{ 1 -  \frac{u^{2n}}{2n}}}    \, , 
\label{defphi}
\end{equation}
 where the angle $\phi$ is defined modulo the
 oscillation period $4 K_n$, with  
\begin{equation}
   K_n   = \sqrt{n} \int_0^1  \frac{{\textrm d}u}{\sqrt{ 1 - u^{2n}}} \, .
\label{nperiod}
\end{equation}
 In terms of the energy-angle  coordinates  $(E, \phi)$,
 the original variables  $(x ,\dot x)$ read as follows: 
 \begin{eqnarray}
          x &=&   E^{1/{2n}} \, {\mathcal S}_n
 \left( \phi  \right) , \label{solnxv2}\\                 
     \dot x &=& (2n)^{\frac{n-1}{2n}} E^{1/2} \,
  {\mathcal S}_n'\left( \phi  \right) \, ,
\label{solnxv3}
 \end{eqnarray}
  where the hyperelliptic function ${\mathcal S}_n$ is defined as 
\cite{abram,byrd}
    \begin{equation}
{\mathcal S}_n(\phi) = Y \,  \leftrightarrow 
  \phi   
  =   \frac{ \sqrt{n}} { (2n)^{1/2n} } \int_0^Y 
   \frac{{\textrm d}u}{\sqrt{ 1 -  \frac{u^{2n}}{2n}}}      \,    .
  \label{hyperelli}
\end{equation}

  The presence of external noise spoils the integrability
 of the dynamical system (\ref{dyn2}) but does
 not preclude the use of $(E,\phi)$ instead of $(x,\dot{x})$  as
 coordinates   in phase space. 
 Introducing  the  auxiliary  variable $\Omega$  defined as 
 \begin{equation}
 \Omega =   (2n)^{ \frac{n+1}{2n} } \,  E^{\frac{1}{2}}  \, ,
 \label{Omega}
\end{equation}
 equation~(\ref{dyn2}) is written  as a system of two coupled
 stochastic differential equations  \cite{philkir2}
   \begin{eqnarray}
     \dot \Omega  &=& n  \, {\mathcal S}_n'(\phi)  \, \xi(t)  
 \label{evolomega}
   \,  ,   \\   
 \dot\phi  &=& \Big ( \frac{\Omega}{ (2n)^{\frac{1}{2n}}} 
\Big )^{\frac{n-1}{n}}
 - \frac{{\mathcal S}_n(\phi)}{\Omega}  \, \xi(t)  \, .
    \label{evolphi}
   \end{eqnarray}
 This system is  rigorously 
 equivalent to the original  problem   (Eq.~\ref{dyn2}) and has been  
  derived without any hypothesis on 
 the  nature of the  driving force   $\xi(t)$  which may even 
 be   a deterministic function  or  may   assume 
  arbitrary statistical properties.
 This  external perturbation  $\xi(t)$ continuously 
 injects energy into the system.
  We have shown  in \cite{philkir2} that,
 when $\xi$ is a Gaussian white  noise or a dichotomous Poisson noise,
 the typical value of   $\Omega$   grows algebraically with time.
 We  have also verified numerically that the same behavior is true for 
 an Ornstein-Uhlenbeck noise. As seen from  Eq.~(\ref{evolphi}), 
 the growth of the phase $\phi$  is faster than that of $\Omega$. 
 Assuming that   in the long time limit   
$\phi$ is uniformly distributed  over  a period
  $[0, 4K_n]$, we obtain, on   averaging
 Eqs.~(\ref{solnxv2},\ref{solnxv3}) over the angle variable,
 the following equipartition relations
\begin{eqnarray}
\langle  E   \rangle   &=&  \frac{n+1}{2n} \, \langle \dot x^2   \rangle \, ,
\label{equipn2}  \\
\langle \dot x^2   \rangle   &=&   \langle  x^{2n}   \rangle  \, .
\label{equipnxv}
\end{eqnarray}
 These  relations  are in excellent agreement with
 the numerical simulations as shown in  Fig. 1.

\begin{figure}
 \hfill
\includegraphics*[width=0.38\textwidth]{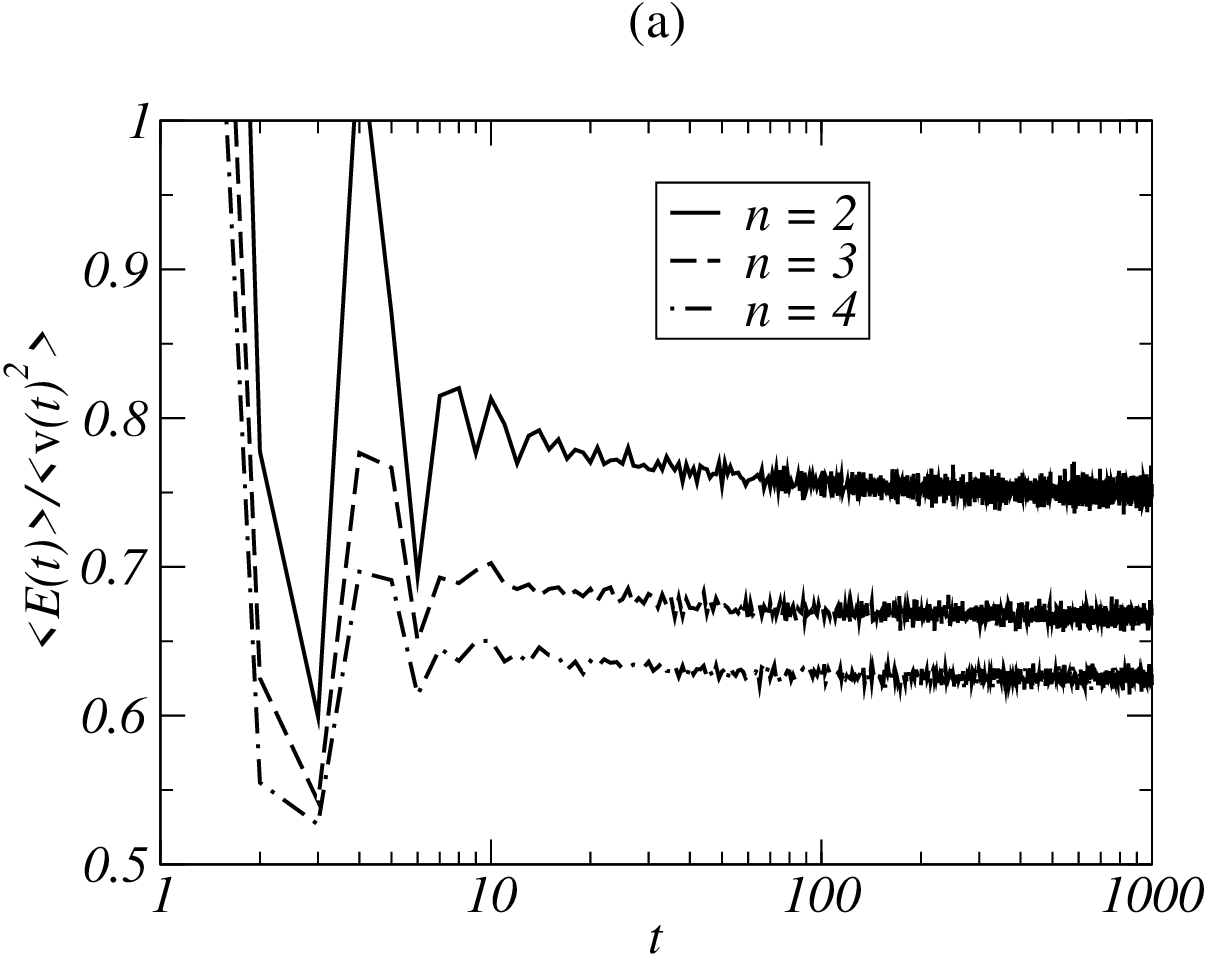}
 \hfill
\includegraphics*[width=0.38\textwidth]{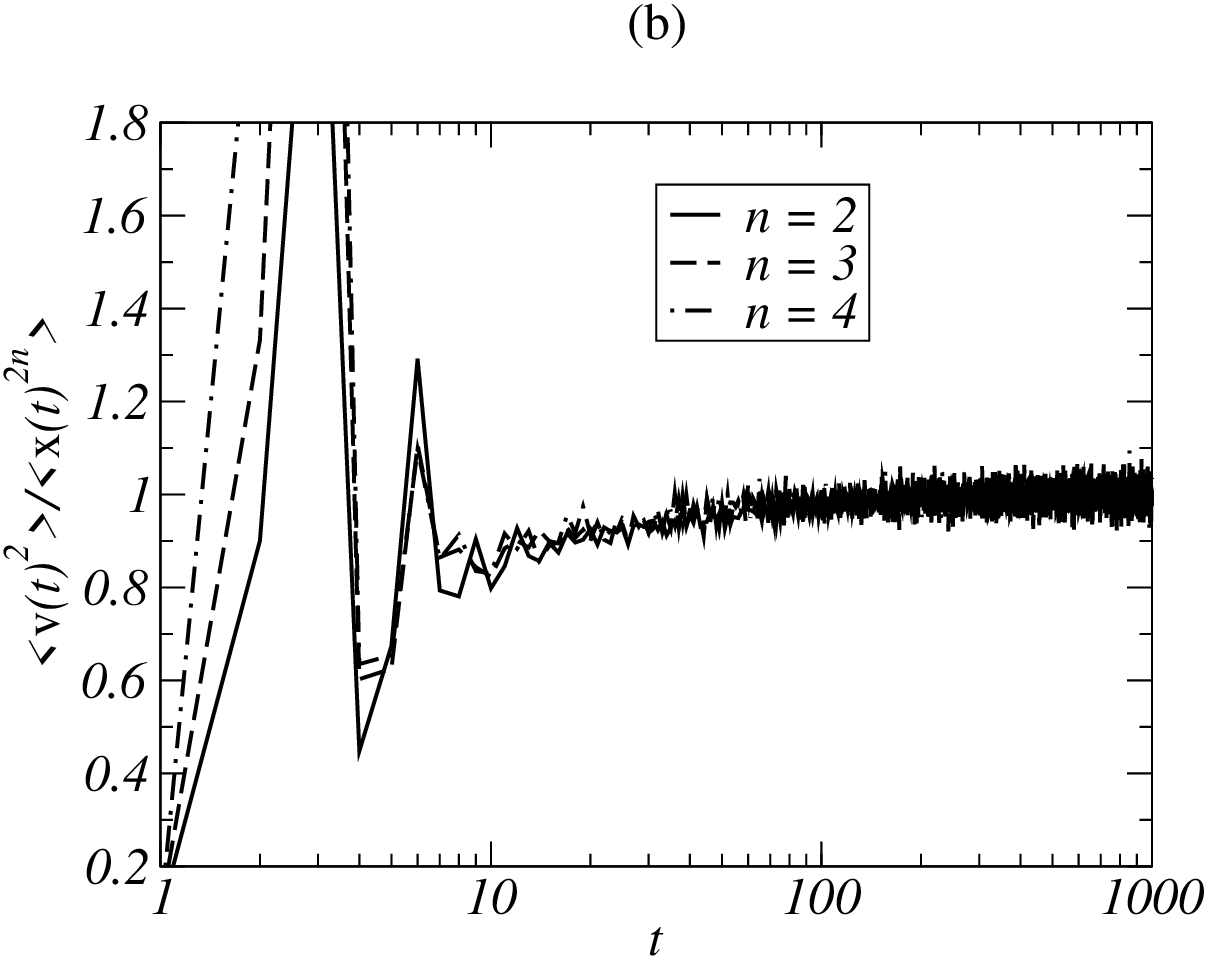}
 \hfill

\caption{Equipartition relations  with  Ornstein-Uhlenbeck noise.
Equations~(\ref{dyn2})~and~(\ref{OU})
  are integrated numerically for ${\mathcal D} = 1$, 
$\tau = 5$, with a timestep $\delta t = 10^{-5}$ 
and averaged over $10^4$ realizations.
Fig.~(a): the first equipartition ratio 
$\langle E(t) \rangle / \langle v(t)^2 \rangle$ is close to
the theoretical value $\frac{n+1}{2n}$  given in  Eq.~(\ref{equipn2}):
$3/4$ for $n=2$; $2/3$ for $n=3$; $5/8$ for $n=4$.
Fig.~(b): the second equipartition ratio 
$\langle v(t)^2 \rangle / \langle x(t)^{2 n} \rangle$ is close to $1$
for $n = 2,3,4$.
}
\label{fig:equip} 
\end{figure}

\subsection{The white noise case}
\label{sec:white}

      In this subsection, we consider  
 $\xi(t)$ to be   a Gaussian white noise
 of zero mean value and  of amplitude ${\mathcal D}$:
\begin{eqnarray}
       \langle \xi(t)  \rangle &=&   0   \, ,\nonumber \\
   \langle \xi(t) \xi(t') \rangle  &=&  {\mathcal D} \, \delta( t - t') .
   \label{defbruitblanc}
 \end{eqnarray}
  As described in \cite{philkir2},
  an  averaging procedure over the angle variable $\phi$
allows  us   to   derive a closed  effective
 Langevin  equation for the slow variable $\Omega$.
 Starting   with  the
 full  Fokker-Planck equation for 
   the evolution of the P.D.F. $P_t(\Omega, \phi)$
  associated  with the system (\ref{evolomega},\ref{evolphi}) and
 integrating  out  the angular variable $\phi$,   we obtain a 
  phase-averaged Fokker-Planck 
 equation for the marginal distribution  $\tilde{P}_t(\Omega)$:
 \begin{equation}
   \partial_t {\tilde P}  =  \frac{  {\tilde {\mathcal D} }}{2} 
\left(  \partial_{\Omega}^2 {\tilde P} - \frac{1}{n} \, \partial_{\Omega}
      \frac{{\tilde P}}{\Omega}  \right)  \, , \,\,\,\,\,\,\hbox{ with  }
 \tilde{{\mathcal  D}} =  \frac{ n^2  (2n)^{\frac{1}{n}} }{ n + 1 }
   {\mathcal  D}    \, . 
 \label{nFPmoy}
\end{equation} 
  The  effective
 Langevin dynamics for the variable $\Omega$  is thus 
\begin{equation}
   \dot\Omega =  \frac{   {\tilde {\mathcal D} } }{2n}
  \, \frac{1}{\Omega}  + {\tilde \xi}(t)  \, ,
\label{nLangeff} 
\end{equation}  
where ${\tilde {\mathcal D} }$   is the amplitude of 
the effective Gaussian white
 noise ${\tilde \xi}(t)$.
 It is readily deduced from  Eq.~(\ref{nLangeff}) that the variable 
  $\Omega$   has  a normal diffusive behavior with time:
\begin{equation}
 \langle  \Omega^2 \rangle  = 
    \frac{  (2n)^{\frac{n+1}{n}}  {\mathcal D}   }{2}  t    \, .
\label{scalingomegablanc}
\end{equation}  
 
The averaged  Fokker-Planck equation (\ref{nFPmoy}) is
  exactly  solvable, leading to  
 the following expression for the energy P.D.F. 
\begin{equation}
\label{pdf2En}
   {\tilde P}_t(E) =
  \frac{1}{  \Gamma \left(\frac{n + 1}{2n}\right)} \,
 \frac{1}{ E } \, \Big( \frac{  (2n)^{\frac{n  + 1}{n}} E}
  { 2\tilde{{\mathcal D}} t } \Big)^{\frac{n  + 1}{2n}}  
   \exp\left\{ 
- \frac{ (2n)^{\frac{n  + 1}{n}} E }{2\tilde{{\mathcal D}} t} \right\} \, ,
\end{equation}
  $\Gamma(.)$   being  the Euler Gamma function \cite{abram}. 
 The asymptotic time dependence  of 
  all moments of the energy, amplitude or velocity  is 
   calculated analytically from
  Eqs.~(\ref{Omega}, \ref{equipn2}, \ref{equipnxv}, and \ref{pdf2En}),
 leading to  the long time scaling behavior of
 dynamical observables as well as   analytical expressions
 for   the  prefactors  (generalized diffusion constants) 
\begin{eqnarray}
 \langle E  \rangle  &=&  \frac{ {\mathcal D} }{2} t 
 \, ,   \nonumber  \\ 
 \langle \dot{x}^2  \rangle  &=&  \frac{n {\mathcal D} }{n+1} t 
  \, , \nonumber    \\
  \langle x^2  \rangle  &=&   
 \frac{ \Gamma\Big( \frac{3}{2n} \Big) }
       { \Gamma\Big( \frac{1}{2n} \Big)}
    \left( \frac{ 2 \; n^2 }{n+1} \; {\mathcal D}  t \right)^ {\frac{1}{n}} 
 \label{scalingwhite} \, .   
\end{eqnarray}
  The physical observables grow algebraically with time  and the
   corresponding anomalous diffusion exponents  depend   only 
 on   the  confining potential at infinity.
 These quantitative
 results   are
 in excellent agreement with numerical simulations  \cite{philkir2}.

\subsection{Scaling behavior for colored noise}
\label{sec:color}

 We now take   $\xi(t)$  to be 
  a   colored  Gaussian noise  with  non-zero
  correlation time $\tau$  and   generated from
 the  Ornstein-Uhlenbeck  equation 
  \begin{equation} 
 \frac{{\textrm d} \xi(t)}{{\textrm d} t} = -\frac{1}{\tau} \xi(t) - 
\frac{1}{\tau} \eta(t)  \, , 
  \label{OU}
\end{equation}
 where $\eta(t)$ is  a Gaussian white noise
 of zero mean value and  of amplitude ${\mathcal D}$. In the stationary limit,
when $t, t' \gg \tau$, we  have:
\begin{equation}
       \langle \xi(t)  \rangle =   0  \;\; \mathrm{and} \;\;
     \langle \xi(t) \xi(t') \rangle  =   
\frac{\mathcal D}{2 \, \tau}   \, {\rm e}^{-|t - t'|/\tau} \,.
   \label{deftau}
 \end{equation} 
 In \cite{philkir2}, we derived  from a self-consistent scaling Ansatz
 that, when the noise has a finite correlation time $\tau$,
 the variable $\Omega$ has a sub-diffusive behavior 
 and grows  as 
 \begin{equation}
   \Omega \sim 
  \Bigg( \frac { {\mathcal D} t } { 2 \, \tau^2}\Bigg)^{\frac{n}{2(2n-1)}} \, .
\label{scalingomegacolor}
\end{equation}  
   From this equation and 
  Eqs.~(\ref{Omega}, \ref{equipn2} and  \ref{equipnxv}), 
  the  following scaling relations are deduced 
\begin{eqnarray}
           E      &\sim&  \Bigg( \frac { {\mathcal D} t } { 2 \, \tau^2}\Bigg)
 ^{\frac{n}{(2n-1)}}  \, , \nonumber     \\
         \dot x  &\sim&  \Bigg( \frac { {\mathcal D} t } { 2 \, \tau^2}\Bigg)
 ^{\frac{n}{2(2n-1)} }  \,  , \nonumber    \\ 
          x      &\sim&  \Bigg( \frac { {\mathcal D} t } { 2 \, \tau^2}\Bigg)
^{\frac{1}{2(2n-1)}}  \, .
\label{scalingcolor}
\end{eqnarray}
 We observe that the exponents associated with colored noise
 are different from those obtained for white noise. We emphasize
 that Eq.~(\ref{scalingcolor})  gives only   qualitative
 scaling relations  that were  obtained only   by heuristic   arguments. In particular,
 the prefactors (which are dimensionless pure numbers) are not
 known at this stage. In the following sections,  we solve this problem analytically: 
 we   calculate 
 the P.D.F. of the anharmonic oscillator subject to
 Ornstein-Uhlenbeck noise and   derive explicit and complete  formulae
 for the moments of the dynamical observables in the long time limit.

\begin{figure}
\centerline{ \includegraphics*[width=0.38\textwidth]{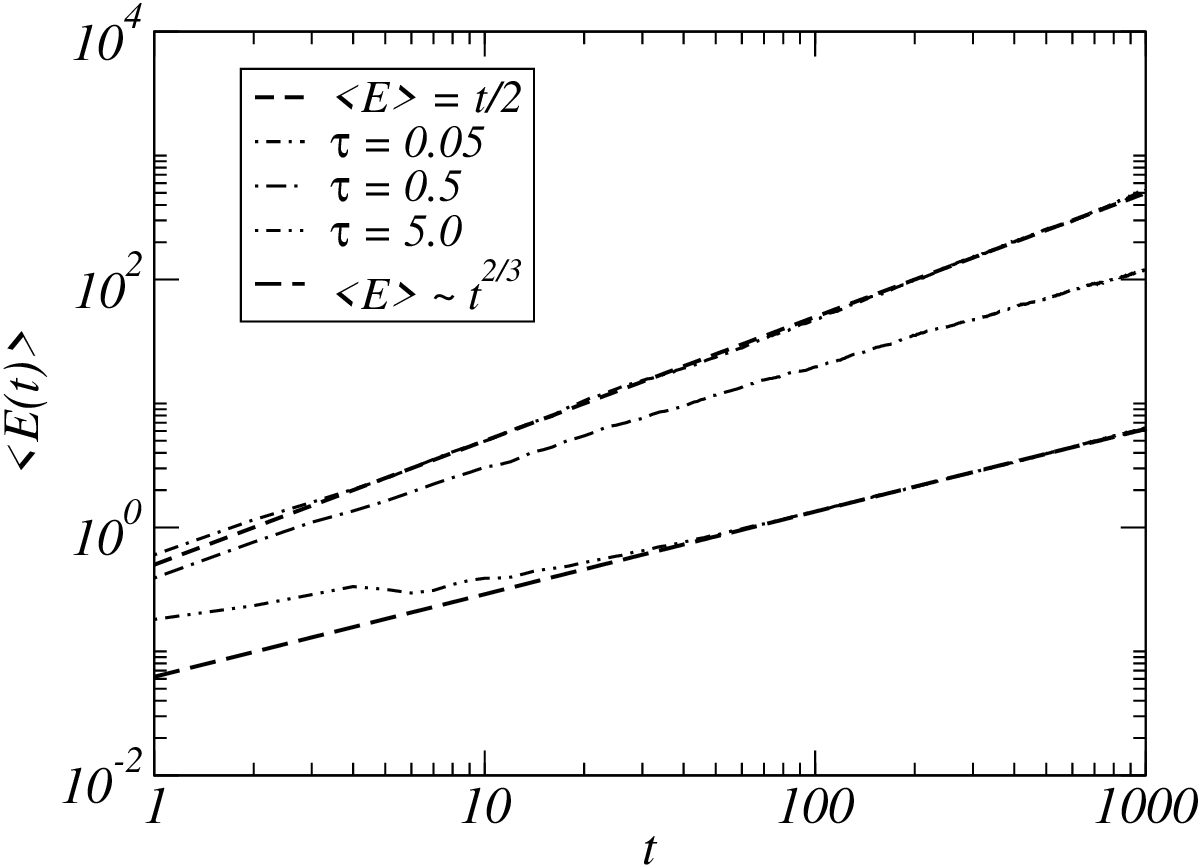}}

    \caption{ Asymptotic  behavior of the
nonlinear oscillator with additive Ornstein-Uhlenbeck noise. 
Eqs.~(\ref{dyn2})~and~(\ref{OU}) 
 are integrated numerically for ${\mathcal D} = 1$, $n = 2$,
with a timestep $\delta t = 10^{-5}$  and averaged over 
$10^3$ realizations. For $\tau = 0.05$, $0.5$ and $5.0$, we plot 
the average $\langle E \rangle$  \emph{vs.}  time $t$.
The dashed lines correspond to the analytical predictions,
$\langle E \rangle = {\mathcal D}\, t/2 $ for white noise 
 (see Eq.~(\ref{scalingwhite}))
and $\langle E  \rangle  = 0.533 \, ( {\mathcal D} \,t / \tau^2 )^{2/3}$
for colored noise with $\tau = 5.0$  (see Eq.~(\ref{tri})). 
}
\label{fig:crossover}
\end{figure}

 The result  of Eq.~(\ref{scalingcolor})   
 is   supported by numerical simulations:  the colored noise scalings
 are observed for large  enough  times, even for  an arbitrarily small 
 correlation time $\tau$ (see Fig.~\ref{fig:crossover}).   The
 crossover between the white noise scalings (\ref{scalingwhite})
 and the colored noise scalings (\ref{scalingcolor}) occurs when 
 the period $T$ of the underlying deterministic oscillator
 (which is a decreasing function of its amplitude) is of
 the order of $\tau$, {\it  i.e.},   for a   typical time 
$t_c \sim {\mathcal{D}}^{-1} \; \tau^{-2n/(n-1)}$.
For $t \ll t_c$,
 the angular period of the system is large as compared  to
 the correlation time of the noise  which, therefore,  acts as if it were
 white;   for $t \gg  t_c$, the noise is highly
 correlated over a period and   is smeared out,
 leading to a slower diffusion. 

 Some interesting  qualitative results have thus already  been obtained.
 But,  a quantitative  study of   the nonlinear
 oscillator (\ref{dyn2}) with  Ornstein-Uhlenbeck noise
 has remained out of reach. The  reason is that 
 the  averaging technique, that applies perfectly  for the white noise,
 leads  to erroneous results when applied to the colored noise.
 We now  explain why  the naive averaging method fails 
 and how to modify it to make it work.

\section{Averaging method for  Ornstein-Uhlenbeck noise}
\label{sec:averaging}

   A   closed  Fokker-Planck equation for
 the probability distribution function $P_t(\Omega, \phi)$ 
   on the system's   phase  space can not be derived 
 when the noise is colored: 
  because  the noise is   correlated in time, the system's dynamics 
 acquires a non-Markovian character and  the evolution
  equation for $P_t(\Omega, \phi)$,  which can be derived  using  functional methods
  \cite{hanggirev1,hanggirev2,hanggirev3,sancho1,sancho2,fox,tsironis},
  involves correlation functions
  at different times. This hierarchy can not be closed
  at any finite order unless truncation  approximations
  are used \cite{hanggirev3,moss}. However, 
 by embedding the initial problem  into  a higher dimensional
   Markovian system, a rigorous description in terms of
  a Fokker-Planck evolution  equation can be formulated. Indeed, 
  the random oscillator  described by Eqs.~(\ref{evolomega},\ref{evolphi})
 together with the Ornstein-Uhlenbeck noise generated by Eq.~(\ref{OU})
 forms a three dimensional stochastic system  driven by a white noise
 $\eta(t)$. The 
 stochastic equations~(\ref{evolomega},\ref{evolphi}  and \ref{OU})
 define   a  random dynamical system
  in the three dimensional space  $(\Omega,\phi,\xi)$, 
    the  three  variables  being  treated on an  equal footing: 
   \begin{eqnarray}
    \dot \Omega  &=&  n  \, {\mathcal S}_n'(\phi)  \, \xi   
 \label{ev1}   \,  ,   \\    
   \dot\phi  &=&  \Big ( \frac{\Omega}{ (2n)^{\frac{1}{2n}}} \Big )^{\frac{n-1}{n}}
 - \frac{{\mathcal S}_n(\phi)}{\Omega}  \, \xi     \, , 
    \label{ev2} \\   
    \dot \xi    &=&  -\frac{1}{\tau} \xi  -  \frac{1}{\tau} \eta(t)   \, ,
  \label{ev3} 
   \end{eqnarray}
   where $\eta(t)$ is a Gaussian white noise of 
   amplitude   ${\mathcal D}$. 
   The  Fokker-Planck equation  for the joint P.D.F.
 $P_t(\Omega,\phi,\xi)$  is then   given by
\begin{equation}
 \frac{ \partial P_t}{\partial t} = - n \frac{\partial }{\partial \Omega}
  \Big( \, {\mathcal S}_n'(\phi) \, \xi \,  P_t \Big) 
  - \frac{\partial }{\partial \phi} \left( \left(  
 \Big ( \frac{\Omega}{ (2n)^{\frac{1}{2n}}} 
\Big )^{\frac{n-1}{n}}
 - \frac{{\mathcal S}_n(\phi)\xi  }{\Omega}    \, \right)  P_t \right)
 + \frac{1}{\tau} \frac{ \partial \xi  P_t}{\partial \xi}
 +\frac{{\mathcal D}}{2 \, \tau^2}\frac{ \partial^2 P_t}{\partial \xi^2}\,  .
\label{FPcolor}
\end{equation}
  If an exact solution  of Eq.~(\ref{FPcolor}) could be found, 
  the P.D.F. on the
 original phase space $(\Omega,\phi)$  will   then  be
 obtained by  integrating the full solution over the variable
 $\xi$. Unfortunately, this does not seem to be feasible.
 However, the evolution of  the angular variable  $\phi$ 
 is fast as compared to that of   $\Omega$ irrespective of
 the correlation time of the noise.  
 Therefore, it is  desirable to average out   $\phi$ from the dynamics
 in order to  obtain an effective Langevin equation  for 
 the slow variable  $\Omega$,  as has been done in  the white noise case.

\subsection{Failure of straightforward  averaging}
\label{sec:straightf}

We now average the  Fokker-Planck
 equation~(\ref{FPcolor})   over the angular variable $\phi$  assuming
 that for  $t \to \infty$   the probability measure 
  for $\phi$ is  uniform over  
 the   interval $[0,4K_n]$. 
  Using  the fact that
 the average of the  derivative  of any function 
 with respect to $\phi$     is zero:
  \begin{equation}
 \overline{ \partial_{\phi}(\ldots)} = 0  \, ,
\end{equation}   
  the  phase-average  of the  Fokker-Planck  equation
(\ref{FPcolor}) governing the evolution
 of the marginal distribution  $\tilde{P}_t(\Omega,\xi)$
 is found to be 
\begin{equation}
 \frac{ \partial \tilde{P}_t}{\partial t} = 
 \frac{1}{\tau} \frac{ \partial \xi  \tilde{P}_t}{\partial \xi}
 +\frac{{\mathcal D}}{2 \, \tau^2}\frac{ \partial^2 \tilde{P}_t}
 {\partial \xi^2}\,  ;
\label{FPavcolor}
\end{equation}
this equation  corresponds to the  following 
  Langevin dynamics for the couple
 of variables $(\Omega,\xi)$
\begin{eqnarray}
     \dot \Omega  &=&  0 \,  \nonumber \\
      \dot \xi   &=& -\frac{1}{\tau} \xi  -   \frac{1}{\tau} \eta(t).
\label{average0}
 \end{eqnarray}
 This result   implies   that   $\Omega$ has a trivial dynamics: it 
 is no more  stochastic  and  is  conserved.  The integration 
  over the angular variable  averages  out the noise
 itself and leads to   conclusions
 that are   blatantly wrong.      
 This failure is due to the  destructive interference  between
 the  fast angular variable  and  the noise:  a simplistic   
 adiabatic elimination of $\phi$   eliminates the noise too. 
 The   failure of this   naive averaging procedure   can also be
  understood from a direct analysis of  
  Eqs.~(\ref{ev1}, \ref{ev2} and \ref{ev3}):   the functions 
  ${\mathcal S}_n(\phi)$  and   ${\mathcal S}_n'(\phi)$ 
  are periodic in $\phi$  with period $4K_n$ 
   and have a  zero mean value. 
  Thus, if we  average  Eqs.~(\ref{ev1} and \ref{ev3})
  over  $\phi$  and   discard the correlations
  between  $\phi$  and $\xi$, the erroneous dynamics~(\ref{average0})
  is obtained.  A correct  averaging scheme  must take into  account the fact
 that the colored noise $\xi$ is not constant  but  varies
  during  the period $T$ of the underlying deterministic oscillator   
 (the  angle $\phi$ covers  the interval  $[0, 4K_n]$ during   $T$);  
  when  $T \ll  \tau$, the variation  
 of  $\xi$ can be estimated from  its time derivative 
 that contains a white noise term  (see Eq.~(\ref{OU})) which 
 will survive the averaging procedure. 

\subsection{A heuristic calculation}
\label{sec:heuristic}

 The preceeding  discussion suggests that we must transform
  the system~(\ref{ev1}, \ref{ev2} and  \ref{ev3}) so as 
  to make   the time  derivative of  $\xi$  appear
  explicitly   in the evolution equation~(\ref{ev1})
 of the slow variable $\Omega$. Before performing this transformation 
  in a systematic manner,
  we explain the heuristics of the method. Taking into account the fact  that
  $\Omega$ grows to infinity with time, let us simplify 
 Eq.~(\ref{ev2}) to   
\begin{equation}
\dot\phi  \simeq \Big ( \frac{\Omega}{ (2n)^{\frac{1}{2n}}} \Big )^{\frac{n-1}{n}}  \,,
\end{equation}
 and   use this relation  to  transform Eq.~(\ref{ev1}) as follows
 \begin{equation}
 \dot \Omega = \frac{ n  {\mathcal S}_n'(\phi)\dot\phi    \, \xi }{ \dot\phi }
  \simeq  \frac{ n  {\mathcal S}_n'(\phi)\dot\phi    \, \xi }
 { \Big ( \frac{\Omega}{ (2n)^{\frac{1}{2n}}} \Big )^{\frac{n-1}{n}} }  \,.
 \label{AppOm}
\end{equation}
Introducing  a  new variable $Y$ defined as
 \begin{equation} 
  Y =  \frac{\Omega^{\frac{2n-1}{n}}}  { (2n)^{\frac{n-1}{2n^2}}} \,,
 \label{defY}
\end{equation}
  we rewrite Eq.~(\ref{AppOm}) as
 \begin{equation}
  \frac{1}{2n-1} \dot  Y =  {\mathcal S}_n'(\phi)\dot\phi    \, \xi 
  = \frac{{\textrm d}  \big ( {\mathcal S}_n(\phi) \xi \big )}{{\textrm d} t}
  - {\mathcal S}_n(\phi) \dot \xi \,. \label{evolY} 
\end{equation}
 Using Ornstein-Uhlenbeck's equation ~(\ref{ev3}), Eq.~(\ref{evolY})
 becomes
 \begin{equation}
 \frac{{\textrm d}}{{\textrm d} t}  
  \big (\frac{Y}{2n-1} - {\mathcal S}_n(\phi) \xi \big )
= \frac{  {\mathcal S}_n(\phi) \xi } {\tau}  +
   {\mathcal S}_n(\phi) \frac{ \eta } {\tau}  \,. \label{evolY2} 
 \end{equation}
 In the long time limit, the term ${\mathcal S}_n(\phi) \xi$     which remains
 of  finite  order  becomes  negligible  as  compared to   $Y$    that 
  grows as a power-law with time.  Besides, on the right hand
  side  of  Eq.~(\ref{evolY2}) we also neglect  
 the colored noise  $\xi$   with respect to 
    the white noise $\eta$.  We thus conclude  that 
 $Y$ has   a normal diffusive behavior   and scales  as
  \begin{equation}
  \langle  Y^2  \rangle  \propto   \frac { {\mathcal D} t } { 2 \, \tau^2} \, .
  \end{equation}
 This scaling  result is equivalent to Eq.~(\ref{scalingomegacolor}) and implies 
  Eq.~(\ref{scalingcolor}). The variable $Y$  undergoes an effective 
  diffusive dynamics  and  plays a role similar to that of  $\Omega$
  in the white noise case.    In order to obtain
 a quantitative agreement between theory and numerics,
 we must set  this  heuristic  calculation  on a sound   basis.
 In particular, some terms   that  were  overlooked 
 in the  sketchy discussion  above   
  contribute to  the long time   behavior of the
 P.D.F. in phase space  and must be taken into account.

\subsection{Coordinate transformations}
\label{sec:coordinate}
 
 A correct  averaging  procedure is  performed  by  defining
 recursively new sets of dynamical variables on the global
 three dimensional phase space   $(\Omega,\phi,\xi)$.
 These new variables  will   embody the correlations
 between  $\phi$ and $\xi$  and   provide a book-keeping device  
 that  transforms the heuristic calculation described  above
  into a systematic perturbative expansion. 
 We show in the  Appendix  B  that a suitable set  of
 variables is given by  $(Z,\phi,\xi)$, where $Z$ is defined as 
 \begin{equation}  
  Z(Y,\phi,\xi)  =  Y - (2n -1) \Big ({\mathcal S}_n(\phi) \xi  
  + \frac {  (2n)^{\frac{n-1}{2n(2n-1)}}   {\mathcal C}_n(\phi) \xi }
  { \tau  Y^{\frac{n-1}{2n -1}}  } 
 +  \frac { {\mathcal S}_n^2(\phi) \xi^2 }{2 Y}  \Big ) \, ;
 \label{defZ}
 \end{equation}
  the variable  $Y$ was  defined in Eq.~(\ref{defY})
 and  the function ${\mathcal C}_n(\phi)$  satisfies
\begin{equation}  
 {\mathcal C}_n'(\phi) = 
 \frac{\textrm{d}\; {\mathcal C}_n(\phi)   }{\textrm{d} \phi}
 = {\mathcal S}_n(\phi) \,\,\, \hbox{ and } \,\,\, \overline{ {\mathcal C}_n(\phi)}
 = \int_0^{4K_n} {\mathcal C}_n(\phi)  {\textrm{d} \phi} = 0 \, ,
 \label{defCn}
\end{equation}
 the overline indicates  an   average over the  angular period.
  
   Equation~(\ref{defZ}) defines a change of dynamical variables 
 between the set $(\Omega,\phi,\xi)$ and  the set 
  $(Z,\phi,\xi)$. The  system~(\ref{ev1}, \ref{ev2} and  \ref{ev3}) 
 must  now be expressed   in terms of the new variables
 and we obtain 
    \begin{eqnarray}
  \frac{1}{2n -1}  \dot Z  &=&  {  J}_Z(Z,\phi,\xi)
 +  {\mathcal D}_Z(Z,\phi,\xi) \frac{ \eta(t)  }{\tau}
\label{evZ}   \,  ,   \\    
   \dot\phi  &=&   { J}_{\phi}(Z,\phi,\xi) \, , 
    \label{evphi} \\   
    \dot \xi    &=&  -\frac{1}{\tau} \xi  -  \frac{1}{\tau} \eta(t)   \, .
  \label{evxi} 
   \end{eqnarray}
 The current  and diffusion functions 
 that appear in Eqs.~(\ref{evZ} and \ref{evphi})  are given by
   \begin{eqnarray} 
 {  J}_Z(Z,\phi,\xi) &=& 
\frac {  (2n)^{\frac{n-1}{2n(2n-1)}}   {\mathcal C}_n(\phi) \xi }
  { \tau^2  Z^{\frac{n-1}{2n -1}}  }  + 
  \frac{ 2 {\mathcal S}_n^2(\phi) + 
 (n-1){\mathcal S}_n'(\phi) {\mathcal C}_n(\phi)}{ \tau Z} \xi^2
     \,  ,  \label{courZ}  \\    
{\mathcal D}_Z(Z,\phi,\xi)  &=&  {\mathcal S}_n(\phi) 
 +  \frac {  (2n)^{\frac{n-1}{2n(2n-1)}}   {\mathcal C}_n(\phi)  }
  { \tau  Z^{\frac{n-1}{2n -1}}  }   +
\frac { {\mathcal S}_n^2(\phi) \xi  }{Z} 
   \,  ,  \label{diffZ} \\    
{    J}_{\phi}(Z,\phi,\xi)  &=&  
 \Big ( \frac{\Omega(Z,\phi,\xi)}{ (2n)^{\frac{1}{2n}}} \Big )^{\frac{n-1}{n}}
 - \frac{{\mathcal S}_n(\phi)}{\Omega(Z,\phi,\xi)}  \, \xi     \, .
\end{eqnarray}
 In  Appendix  B,  we show that 
 the stochastic dynamical  system~(\ref{evZ}, \ref{evphi} and  \ref{evxi}) 
 is obtained  from  the original equations~(\ref{ev1}, \ref{ev2} and  \ref{ev3}) 
  by retaining all  contributions  up  the   order ${\mathcal O}(Z^{-1})$.
  The crucial difference 
 between the two systems appears when  Eq.~(\ref{ev1}) is compared 
 with  Eq.~(\ref{evZ}): the evolution of $Z$ contains an explicit white
 noise contribution that embodies the variations of the colored  noise  $\xi$
 during a time scale shorter than the correlation time $\tau$. This white
 noise  term will survive the averaging process and  will allow us to derive
 a non-trivial  Langevin  dynamics  for $Z$.

\subsection{Averaged  Langevin system}
\label{sec:averagedLang}

 As for the white noise problem, we shall  eliminate 
 the fast angular variable in  
the  Fokker-Planck equation  for the  P.D.F.
 $\Pi_t(Z,\phi,\xi)$  associated  with 
 Eqs.~(\ref{evZ}, \ref{evphi} and  \ref{evxi}). This Fokker-Planck equation
  is  given by
\begin{eqnarray}
  \frac{ \partial \Pi_t}{\partial t} =  &&
  - (2n -1) \frac{\partial }{\partial Z}
  \left( { J}_Z    \,  \Pi_t \right) 
  - \frac{\partial }{\partial \phi}
  \left(  { J}_{\phi}  \,  \Pi_t \right)  
 + \frac{1}{\tau} \frac{ \partial \xi  \Pi_t}{\partial \xi} +
    \nonumber  \\
  &&  \frac{{\mathcal D}}{2 \, \tau^2} \Bigg\{
 (2n-1)^2 
\frac{\partial }{\partial Z} {\mathcal D}_Z \frac{\partial }{\partial Z}
  ({\mathcal D}_Z  \Pi_t)  -  (2n-1) \frac{\partial^2 }{\partial \xi \partial Z  }
  ({\mathcal D}_Z  \Pi_t)  
 -  (2n-1)   \frac{\partial }{\partial Z}
 {\mathcal D}_Z \frac{ \partial  \Pi_t}{\partial \xi} + 
  \frac{ \partial^2 \Pi_t}{\partial \xi^2}   \Bigg\}\,     .
\label{FPZ}
\end{eqnarray}
  Integrating  out   the  fast  variable $\phi$, we obtain 
    the following averaged Fokker-Planck equation
  that describes the evolution of the marginal P.D.F.
  $\tilde\Pi_t(Z,\xi)$   (the details of the calculations   are 
  explained in Appendix C) 
 \begin{eqnarray}
 &&  \frac{ \partial \tilde\Pi_t}{\partial t} = 
 - (2n -1) \mu  \frac{\partial }{\partial Z}
 \Bigg( \frac{ (3 -n)\xi^2  } 
 {\tau Z}  \tilde\Pi_t  \Bigg)
  + \frac{1}{\tau} \frac{ \partial \xi  \tilde\Pi_t}{\partial \xi} + 
    \\ &&
  \frac{{\mathcal D }}{2 \, \tau^2} \Bigg\{
 (2n-1)^2 \mu 
\frac{\partial^2 \tilde\Pi_t  }{\partial Z^2}
  -   (2n-1)   \mu \frac{\partial^2 }
 {\partial \xi \partial Z   }  \Bigg(   \frac{\xi \tilde\Pi_t }{Z} \Bigg)   
  -  (2n-1)  \mu \frac{\partial }{\partial Z} \Bigg(   \frac{\xi}{Z} 
  \frac{ \partial \tilde\Pi_t}{\partial \xi} \Bigg) 
  +  \frac{ \partial^2 \tilde\Pi_t}{\partial \xi^2}   \Bigg\}\, ,  
  \label{FPZav}   
\end{eqnarray}
 where the parameter $\mu$, calculated in Appendix A,  is given by 
\begin{equation}
  \mu  = \overline{ {\mathcal S}_n^2 } =  (2n)^{ \frac{1}{n}} 
    \frac{ \Gamma\Big( \frac{3}{2n} \Big)\Gamma\Big( \frac{n+1}{2n} \Big)     }
       { \Gamma\Big( \frac{1}{2n} \Big)\Gamma\Big( \frac{n+3}{2n} \Big)   }  \, .
 \label{defmu}
 \end{equation}

   Let us now  consider the following  Langevin system for 
   the variables $Z$ and $\xi$: 
\begin{eqnarray}
 \dot Z  &=&  (2n -1) \mu \frac{ (3 -n)\xi^2  } {\tau Z} 
 + (2n -1)   \frac{ \sqrt{\mu} }{\tau} \tilde\eta(t)
 +    (2n -1) \mu \frac{\xi}{\tau Z} \eta(t)   \, , 
  \label{avZ}  \\
 \dot \xi    &=&  -\frac{1}{\tau} \xi  -  \frac{1}{\tau} \eta(t)   \, ,  
  \label{avxi} 
\end{eqnarray}
 where $\tilde\eta(t)$ and $\eta(t)$ are two independent Gaussian
 white noises of amplitude ${\mathcal D}$.  Writing 
 the  Fokker-Planck equation  related to this  stochastic
 system~(\ref{avZ}, \ref{avxi}) 
  we retrieve  the  averaged Fokker-Planck equation~(\ref{FPZav})  up to 
 terms of the order  ${\mathcal O}(Z^{-2})$  (recall that the stochastic 
  Eqs.~(\ref{evZ}, \ref{evphi} and  \ref{evxi}) are 
 equivalent to  Eqs.~(\ref{ev1}, \ref{ev2} and  \ref{ev3}) 
  at  order  ${\mathcal O}(Z^{-1})$). We have thus found 
   the  effective  stochastic Langevin equation~(\ref{avZ})   
 for the slow variable  $Z$ coupled with the Ornstein-Uhlenbeck  noise  $\xi$.
 This dynamics is however  not exactly solvable and we  need 
  one last transformation before  deriving  explicit
 analytical results.  In terms of the  new variable
  \begin{equation}
    Z_1  = Z +   (2n -1) \mu \frac{ \xi^2  } { 2  Z}   \, , 
  \label{defZ2}
  \end{equation}
  Eq.~(\ref{avZ})  becomes 
 \begin{equation}
  \dot  Z_1  = \dot Z  + (2n -1) \mu \frac{ \xi  \dot \xi  } { 2  Z}
 =  (2n -1) \mu \frac{ (2 -n)\xi^2  } {\tau Z_1} 
 + (2n -1)   \frac{ \sqrt{\mu} }{\tau} \tilde\eta(t)   \, ,  
\label{avZ2}
\end{equation}
 where we have neglected all  terms of order strictly less than  ${\mathcal O}(Z^{-1})$.
   Recalling that 
 $\xi^2$ has a  finite mean value  equal to ${\mathcal D}/2\tau$,
 we rewrite the Langevin equation~(\ref{avZ2}) for  $Z_1$
 as follows
 \begin{equation}
  \dot  Z_1  =   (2n -1)(2 -n)\frac{\mu {\mathcal D}}{2\tau^2} \frac{1}{Z_1}
     + (2n -1)   \frac{ \sqrt{\mu} }{\tau} \tilde\eta(t)   
 +  (2n -1)  (2 -n)\frac{ \mu } {\tau Z_1}
   \Big( \xi^2 - \langle \xi^2 \rangle  \Big)  \, . 
\label{LangZ2}
\end{equation}
 This  equation contains two  independent
  noise sources: a white noise contribution,  $\tilde\eta(t)$, 
 and a non-Gaussian colored noise term, $(\xi^2 - \langle \xi^2 \rangle)$,
 of zero mean and of finite variance   $ {\mathcal D}^2/2\tau^2$.
This colored noise    is multiplied  by a  factor
 proportional to $1/Z_1$ and therefore its effect becomes negligible
 as compared to that  of  $\tilde\eta(t)$ when $t \to \infty$. We thus
 simplify  Eq.~(\ref{LangZ2}) to
 \begin{equation}
  \dot  Z_1  =   (2n -1)(2 -n)\frac{\mu {\mathcal D}}{2\tau^2} \frac{1}{Z_1}
     + (2n -1)   \frac{ \sqrt{\mu} }{\tau} \tilde\eta(t)     \, . 
\label{LangeffZ2}
\end{equation}
 We have thus  obtained  an effective white noise
 Langevin dynamics for the variable   $Z_1$. This equation
 plays the same role    for the  nonlinear oscillator subject
  to   a  colored Ornstein-Uhlenbeck
 noise  as that  played by Eq.~(\ref{nLangeff}) for  
 white noise. The mathematical structure of 
 the two equations~(\ref{nLangeff}) and~(\ref{LangeffZ2}) is the same
 and they   differ  only  by the constant prefactors that embody
 the  specific parameters of  each  problem.

 \section{Analytical results  in the long time limit}
\label{sec:results}

     In this section, we solve the effective dynamics for the 
 variable   $Z_1$  and deduce  statistical results
 for the long time behavior of the energy, the amplitude
 and the velocity of the nonlinear oscillator driven by an Ornstein-Uhlenbeck
 process. We shall not only  retrieve  the scalings 
 predicted in Eq.~(\ref{scalingcolor}) but also derive explicit  formulae for the 
 prefactors  and  for  the skewness and the flatness of the energy P.D.F.
 All these analytical results   compare favorably  with numerical simulations.
 Finally, we show that  our analysis   can be extended to include  a weak dissipation 
 term  in the dynamics.  

\subsection{Calculation of the P.D.F.}

  The Fokker-Planck equation   associated with 
 the  effective Langevin dynamics~(\ref{LangeffZ2})  of $Z_1$  is exactly solvable
  (using  the method of \cite{philkir1})   and we obtain  the
 following expression for the asymptotic P.D.F.
 \begin{equation}
  P_t(Z_1)  = \frac{ 2}{  \Gamma \left(\frac{n + 1}{2(2n-1)}\right)}
 \Big(  \frac{\tau^2   }{2 (2n -1)^2 \mu {\mathcal D} t } 
  \Big)^{\frac{n + 1}{2(2n-1)}}  Z_1^{\frac{ 2 -n}{2n -1}} 
 \exp\Big( -\frac{ \tau^2 Z_1^2}{ 2 (2n-1)^2\mu {\mathcal D} t }\Big)    \, . 
 \label{PDFZ2}
 \end{equation}

 Expressing  the energy $E$ as a function of $Z_1$ with the help of 
 Eqs.~(\ref{Omega}, \ref{defY}, \ref{defZ} and \ref{defZ2}), 
 we obtain at the  leading  order 
\begin{equation}
    E  = \Bigg( \frac{Z_1}{2n}\Bigg)^{\frac{2n}{2n-1}} \,\,\, \hbox{ and thus  } \,\,\,
   {\textrm d}E =  \frac{1}{2n-1}  \Bigg( \frac{Z_1}{2n}\Bigg)^{\frac{1}{2n-1}}
    {\textrm d}Z_1 \, .
 \end{equation}
 From  Eq.~(\ref{PDFZ2}),  we   deduce the  asymptotic expression 
 of the  P.D.F. of the energy 
 \begin{equation}
    P_t(E)  = \frac{\frac{ 2n -1}{n}}{  \Gamma \left(\frac{n + 1}{4n-2}\right)}
   \Big( \frac{ 2n^2 \tau^2  }   { (2n-1)^2\mu {\mathcal D} t }
   \Big)^{\frac{n + 1}{4n-2}} 
    E^{\frac{1 -n}{2n}} 
  \exp\Big( -\frac{ 2n^2 \tau^2 E^{\frac{2n-1}{n}} }
  { (2n-1)^2\mu {\mathcal D} t }\Big)    \, . 
 \label{PDFE}
 \end{equation}
 The statistical  behavior of the position, velocity and energy of the system
 in the long time limit can now be derived by using the P.D.F.~(\ref{PDFE})
 and Eqs.~(\ref{solnxv2} and \ref{solnxv3}): 
\begin{eqnarray}
  \langle E \rangle &=&
  \frac{ \Gamma \left(\frac{3n + 1}{4n-2}\right)}
       {\Gamma \left(\frac{n + 1}{4n-2}\right)}
\Big(\frac { (2n-1)^2\mu {\mathcal D} t } {2n^2 \tau^2 }\Big)^{\frac{n}{2n-1}} 
   \label{moyE} \, , \\
    \langle \dot x^2 \rangle &=&   \frac{2n}{n+1}  \langle E \rangle 
  \label{moyv2} \, , \\
        \langle  x^2 \rangle &=&   \mu  \langle E^\frac{1}{n} \rangle 
 = \frac{ \Gamma \left(\frac{n + 3}{4n-2}\right)}
       {\Gamma \left(\frac{n + 1}{4n-2}\right)}
\Big(\frac { (2n-1)^2\mu^{2n} {\mathcal D} t } {2n^2 \tau^2 }\Big)^{\frac{1}{2n-1}} 
  \label{moyx2} \, . 
\end{eqnarray}
 In the last equality   we used   $\mu  = \overline{ {\mathcal S}_n^2 }$, 
  according to Eq.~(\ref{defmu}).
  In particular,  we find  \hfill\break
For $ n =2 $, 
\begin{equation}
  \langle E  \rangle  = 0.533 \,
 \Big( \frac {\mathcal{D} \,t } {\tau^2 } \Big)^{2/3} \,  , 
  \,   \langle \dot{x}^2  \rangle  = 0.711  
   \,\Big( \frac {\mathcal{D} \,t } {\tau^2 } \Big)^{2/3}  \,   ,
   \,   \langle x^2  \rangle     =  0.587   \, 
\Big( \frac {\mathcal{D} \,t } {\tau^2 } \Big)^{1/3}     \, .
 \label{tri}   
\end{equation}
For $ n =3 $,
\begin{equation}
    \langle E  \rangle  = 0.474 \,
\Big( \frac {\mathcal{D} \,t } {\tau^2 } \Big)^{3/5} \,  , 
  \,   \langle \dot{x}^2  \rangle  =  0.711   
 \,\Big( \frac {\mathcal{D} \,t } {\tau^2 } \Big)^{3/5}  \,   ,
   \,   \langle x^2  \rangle     =  0.535  \, 
\Big( \frac {\mathcal{D} \,t } {\tau^2 } \Big)^{1/5}     \,  .
 \label{penta}   
\end{equation}
For $ n =4 $,
\begin{equation}
  \langle E  \rangle  = 0.436  \,
\Big( \frac {\mathcal{D} \,t } {\tau^2 } \Big)^{4/7} \,  ,
  \,  \langle \dot{x}^2  \rangle  = 0.698    \,
\Big( \frac {\mathcal{D} \,t } {\tau^2 } \Big)^{4/7}  \,    ,  
   \,   \langle x^2  \rangle  = 0.500    \,
\Big( \frac {\mathcal{D} \,t } {\tau^2 } \Big)^{1/7}      \,  . 
\label{hepta}
\end{equation}

    From  Eq.~(\ref{PDFZ2}),  the skewness and the flatness of the
 energy P.D.F. can also be calculated 
\begin{eqnarray}
S(E) &=& \frac{\langle E^3 \rangle}{\langle E^2 \rangle^{3/2}}
=   \frac{ \Gamma \left(\frac{7n + 1}{4n-2}\right)
 \Gamma \left(\frac{n + 1}{4n-2}\right)^{\frac{1}{2}}  }
 {\Gamma \left(\frac{5n + 1}{4n-2}\right)^{\frac{3}{2}}     } 
  \,   ,   \label{skewE}\\
  F(E) &=& \frac{\langle E^4 \rangle}{\langle E^2 \rangle^2}
=  \frac{ \Gamma \left(\frac{9n + 1}{4n-2}\right)
 \Gamma \left(\frac{n + 1}{4n-2}\right)  } 
 {\Gamma \left(\frac{5n + 1}{4n-2}\right)^2   } 
 \, .   \label{flatE}
\end{eqnarray}
The skewness is approximatively equal to 1.95 for 
 $ n =2, 3 $ and 4 and its limiting value for  $n \to
 \infty$ is given by 
$ \frac{\Gamma(\frac{7}{4}) \; \Gamma(\frac{1}{4})}{\Gamma(\frac{5}{4})^{3/2}}
 =  \frac{6\sqrt{2} \pi}{\Gamma(\frac{1}{4})^2}
 \simeq  2.03\ldots$ The flatness is 
approximatively equal to 4.7 for 
 $ n =2, 3 $ and 4 and its limiting value for  $n \to
 \infty$ is equal to 5.

\begin{figure}
 \hfill
 \includegraphics*[width=0.38\textwidth]{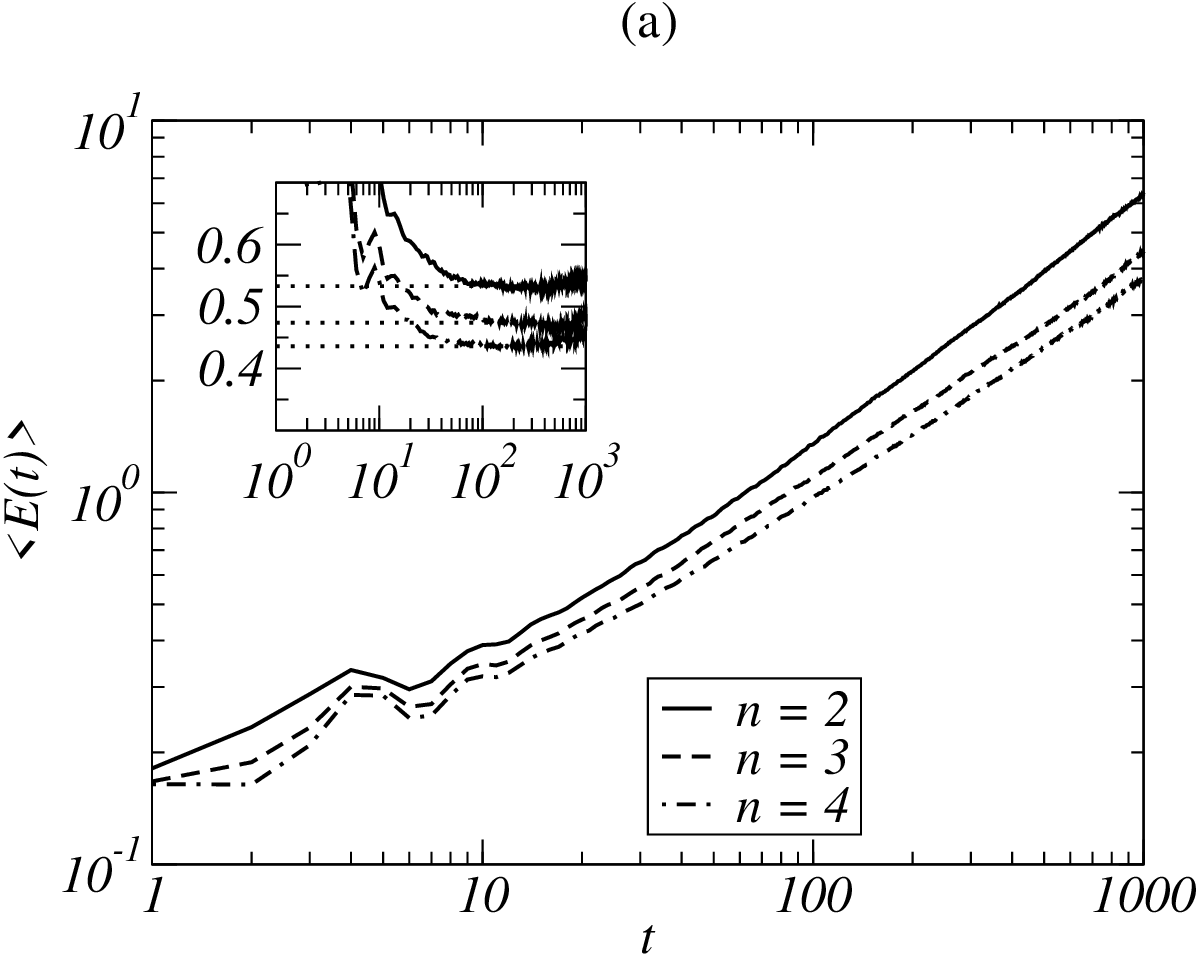}
 \hfill
 \includegraphics*[width=0.34\textwidth]{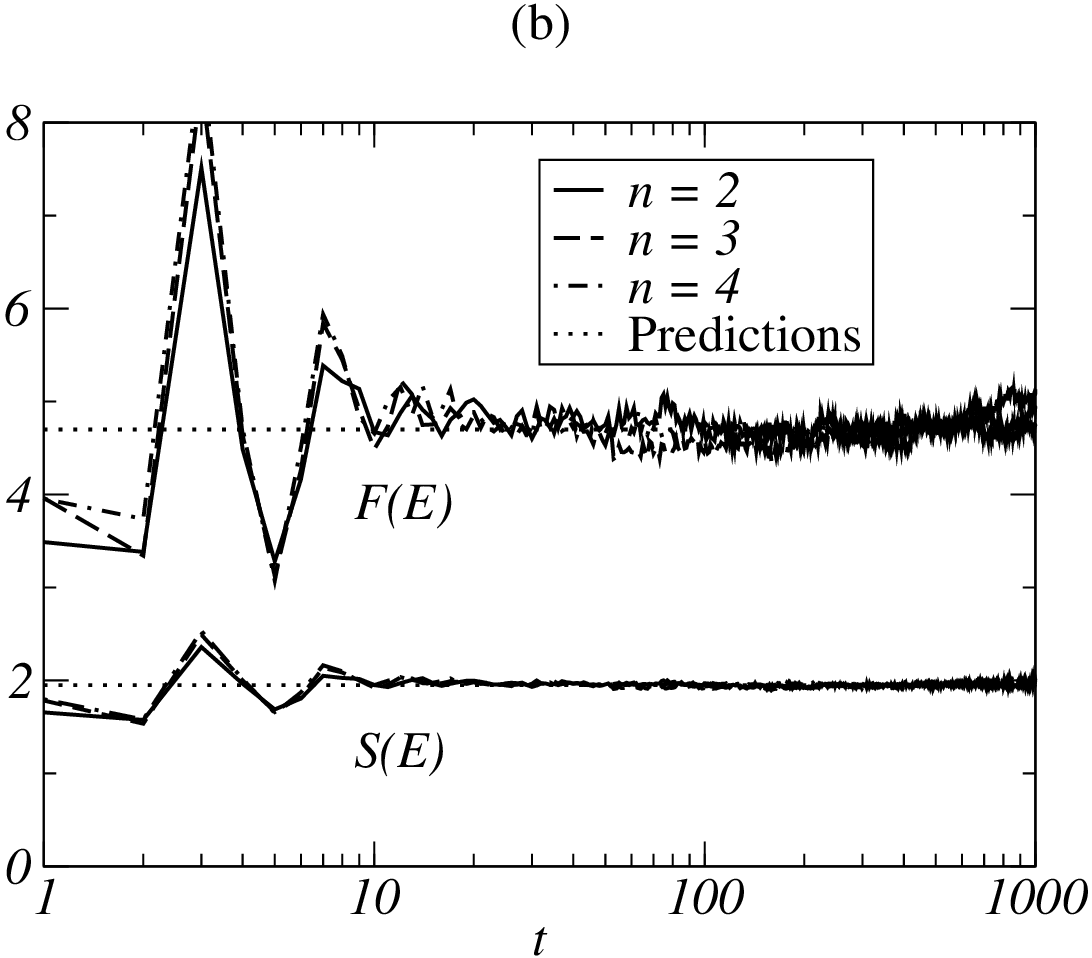}
 \hfill

\bigskip
\centerline{ \includegraphics*[width=0.38\textwidth]{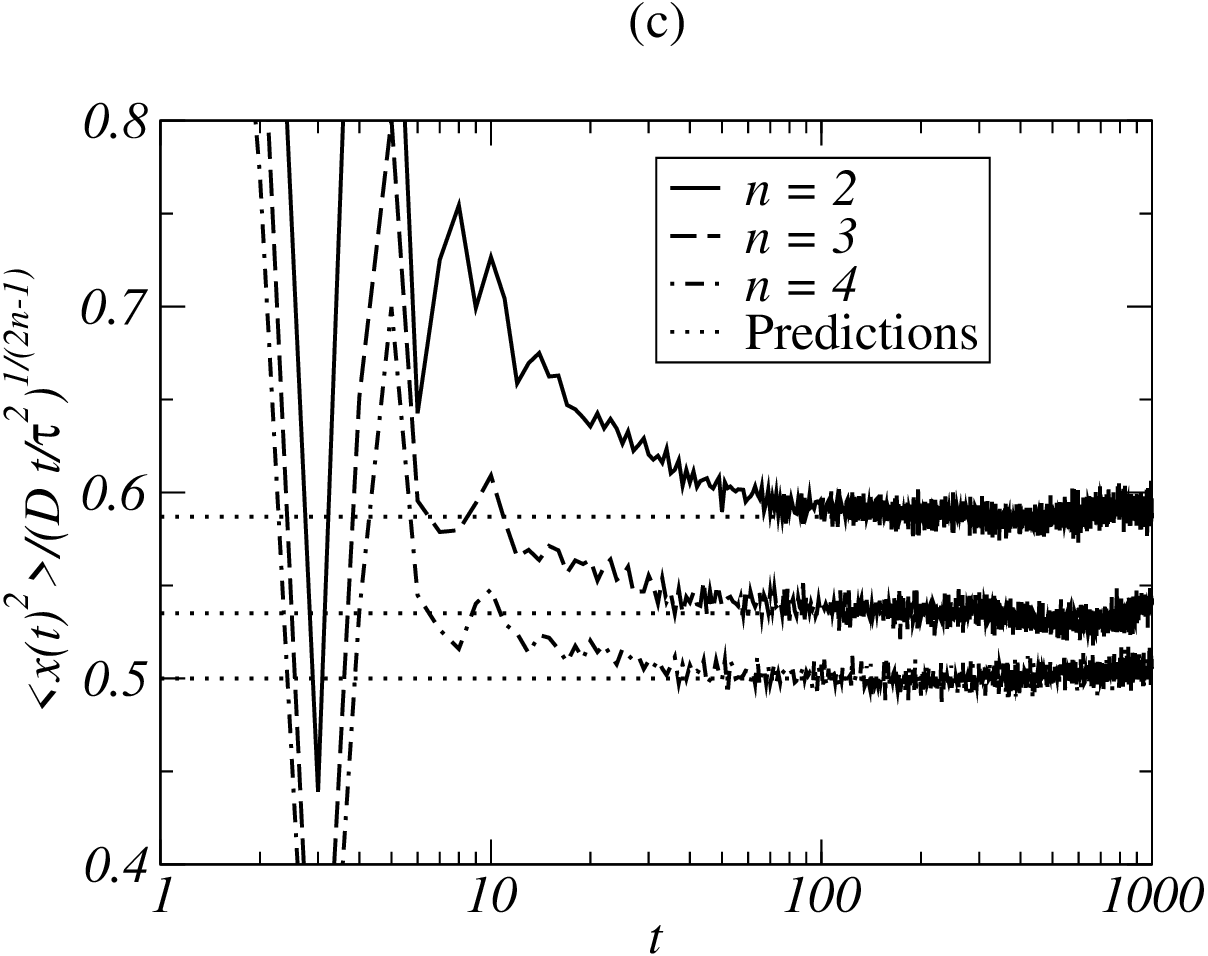}}

    \caption{ Asymptotic  behavior of the
nonlinear oscillator with additive Ornstein-Uhlenbeck noise. 
Eqs.~(\ref{dyn2})~and~(\ref{OU}) 
 are integrated numerically for ${\mathcal D} = 1$, 
$\tau = 5$, with a timestep $\delta t = 10^{-5}$  
and averaged over $10^4$ realizations. For $n = 2,3,4$, we plot: 
$(a)$ the average $\langle E \rangle$ and the ratio 
$\langle E \rangle/(\mathcal{D} \,t/\tau^2)^{n/(2n-1)}$ (inset);
$(b)$ the skewness and flatness factors of $E$ \emph{vs.}  time $t$
$(c)$ the ratio $\langle x^2 \rangle/(\mathcal{D} \,t/\tau^2)^{1/(2n-1)}$.
The dotted lines in the figures correspond to the analytical predictions,
Eqs.~(\ref{moyE}),~(\ref{moyx2}),~(\ref{skewE})~and~(\ref{flatE}).
}
\label{fig:scaling}
\end{figure}

 The  analytical results of   Eqs.~(\ref{tri}-\ref{flatE})  are
 compared with numerical simulations in Fig.~\ref{fig:scaling}.
 We observe that the agreement is remarkable in the long time limit. 
 
\subsection{Extension to the case  of a weakly dissipative system}
\label{sec:extension}

   We now  introduce a linear friction  term in
 the system  with dissipation rate
 $\gamma$. Equation~(\ref{PDFZ2}) then  becomes
  \begin{equation}
   \frac{\textrm{d}^2 }{\textrm{d} t^2}x(t) 
 + \gamma  \frac{\textrm{d}}{\textrm{d} t}x(t) 
  + x(t)^{2n-1}  =  \xi(t)  \,.
 \label{dissdyn2}
\end{equation}
 In terms of the  $(\Omega, \phi)$ variables  defined in
 Eqs.~(\ref{solnxv2}, \ref{solnxv3} and  \ref{Omega}),
 Eq.~(\ref{dissdyn2}) becomes 
 \begin{eqnarray}
     \dot \Omega  &=& -  \frac{ n \gamma{{\mathcal S}_n'}^2(\phi)  }
  { (2n)^{\frac{1}{n}} }  \Omega  + n  \, {\mathcal S}_n'(\phi)  \, \xi(t)  
 \label{dissomega}
   \,  ,   \\   
  \dot\phi  &=& \frac{ \gamma {\mathcal S}_n(\phi){\mathcal S}_n'(\phi) }
  { (2n)^{\frac{1}{n}} }   +  
  \Big ( \frac{\Omega}{ (2n)^{\frac{1}{2n}}} \Big )^{\frac{n-1}{n}}
  - \frac{{\mathcal S}_n(\phi)}{\Omega}  \, \xi(t)  \, .
     \label{dissphi}
   \end{eqnarray}
 We want  to average  out 
  the angular variable from  this system; this is possible 
  only  if  the evolution of $\phi$ is  fast as compared to that  
  of the energy~: this condition 
 is satisfied  when the dissipation rate
 is vanishingly small so that it has a negligible effect
 during a period of the underlying deterministic oscillator.  
 We  proceed as  in Section~\ref{sec:averaging}  
 and obtain  after adiabatic elimination of the angle 
 the following white noise  Langevin equation for the effective
 variable $Z_1$ [defined in Eq.~(\ref{defZ2})]
 \begin{equation}
  \dot  Z_1  =  -  \frac{(2n -1)\gamma}{n+1} Z_1  + 
   (2n -1)(2 -n)\frac{\mu {\mathcal D}}{2\tau^2} \frac{1}{Z_1}
     + (2n -1)   \frac{ \sqrt{\mu} }{\tau} \tilde\eta(t)     \, . 
\label{LangdissZ2}
\end{equation}
 To derive this equation we have assumed that  $\gamma \tau \ll 1 ,$
 and   have used the  identity
 $ \overline{ {{\mathcal S}_n'}^2 } = \frac{ (2n)^{\frac{1}{n}} }{n+1} \,  $
 (derived in Appendix \ref{sec:app0}). When $ t \to \infty$,
 the P.D.F. of  $Z_1$  tends to  a well defined
 stationary  limit   $ P_{\textrm{stat}}(Z_1)$  
 which can be explicitly calculated by solving the stationary
 Fokker-Planck equation  associated with  Eq.~(\ref{LangdissZ2}).
 Reverting to the energy variable $E$, we obtain the stationary
 distribution function of $E$
 \begin{equation}
    P_{\textrm{stat}}(E)  = 
 \frac{\frac{ 2n -1}{n}}{  \Gamma \left(\frac{n + 1}{4n-2}\right)}
   \Big( \frac{ 4n^2 \gamma \tau^2  }   { (2n-1)(n+1)\mu {\mathcal D}  }
   \Big)^{\frac{n + 1}{4n-2}} 
    E^{\frac{1 -n}{2n}} 
  \exp\Big( -\frac{ 4n^2 \gamma \tau^2 E^{\frac{2n-1}{n}} }
  { (2n-1)(n+1) \mu {\mathcal D}  }\Big)    \, . 
 \label{PDFstatE}
 \end{equation}
 We have thus derived, in the limit of vanishingly small dissipation, 
 an analytical  expression  for   the stationary  probability 
 distribution function of the energy of a non-linear oscillator in
 presence of an additive Ornstein-Uhlenbeck noise. 
  In contrast to the case of white noise,   this stationary
  P.D.F.  is not the canonical Gibbs-Boltzmann distribution.

 The statistical  behavior of the position, velocity and energy of the system
 in the long time limit can now be derived by using the P.D.F.~(\ref{PDFstatE})
 and Eqs.~(\ref{solnxv2} and \ref{solnxv3}): 
\begin{eqnarray}
  \langle E \rangle &=&
  \frac{ \Gamma \left(\frac{3n + 1}{4n-2}\right)}
       {\Gamma \left(\frac{n + 1}{4n-2}\right)}
\Big(\frac {(2n-1)(n+1)\mu {\mathcal D}} {4n^2\gamma \tau^2 }\Big)^{\frac{n}{2n-1}} 
   \label{moystatE} \, , \\
    \langle \dot x^2 \rangle &=&   \frac{2n}{n+1}  \langle E \rangle 
  \label{moystatv2} \, , \\
        \langle  x^2 \rangle &=&   \mu  \langle E^\frac{1}{n} \rangle 
 = \frac{ \Gamma \left(\frac{n + 3}{4n-2}\right)}
       {\Gamma \left(\frac{n + 1}{4n-2}\right)}
\Big(\frac { (2n-1)(n+1)\mu^{2n} {\mathcal D}  }
  {4n^2\gamma \tau^2 }\Big)^{\frac{1}{2n-1}} 
  \label{moystatx2} \, . 
\end{eqnarray}
  We observe that  Eqs.~(\ref{PDFstatE}  to \ref{moystatx2})
  can be obtained   from  Eqs.~(\ref{PDFE}  to \ref{moyx2})   
 by making the substitution 
$\frac{1}{2t} = \frac{2n-1}{n+1}\gamma .$
 
\begin{figure}
 \hfill
 \includegraphics*[width=0.38\textwidth]{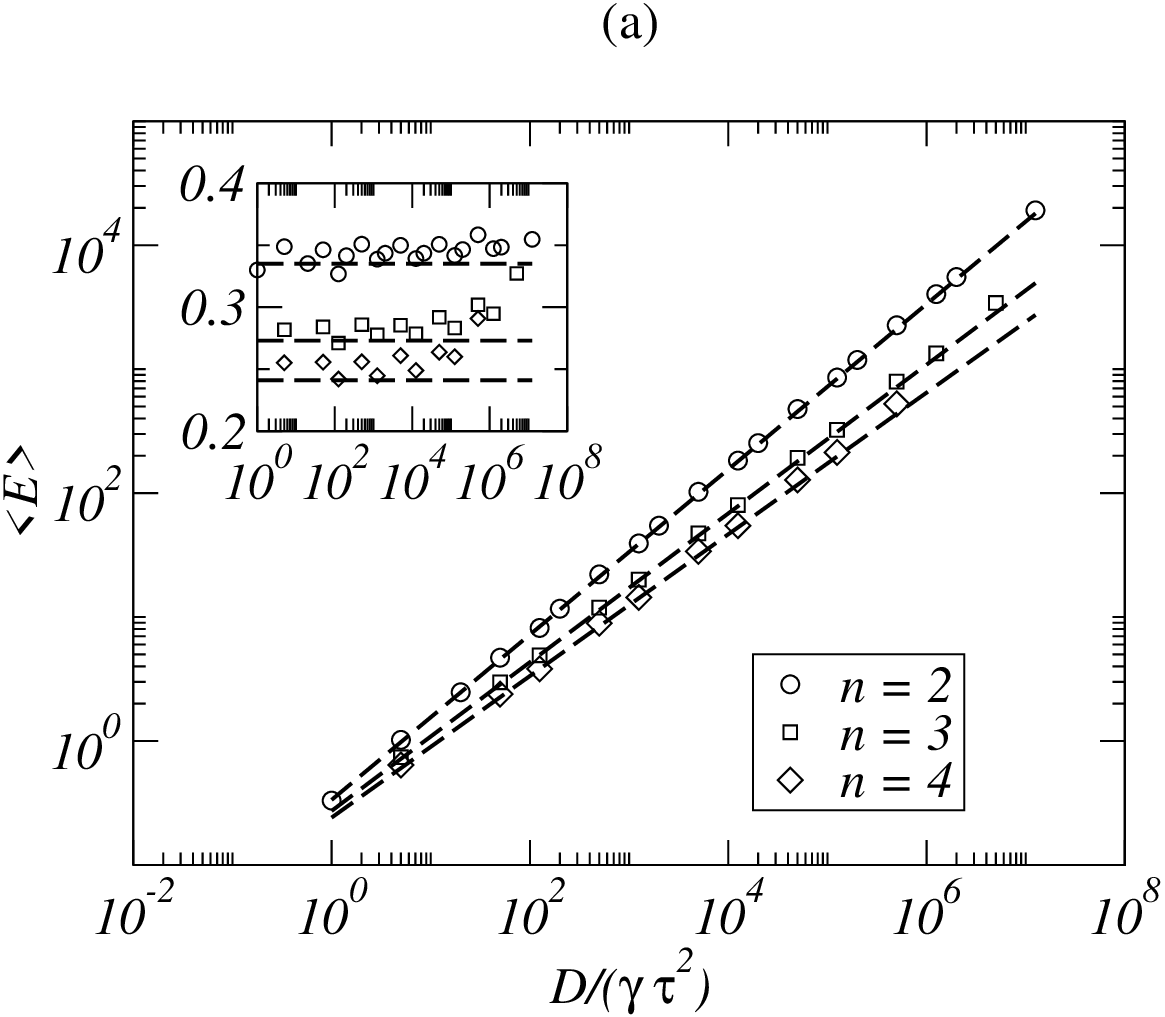}
 \hfill
 \includegraphics*[width=0.38\textwidth]{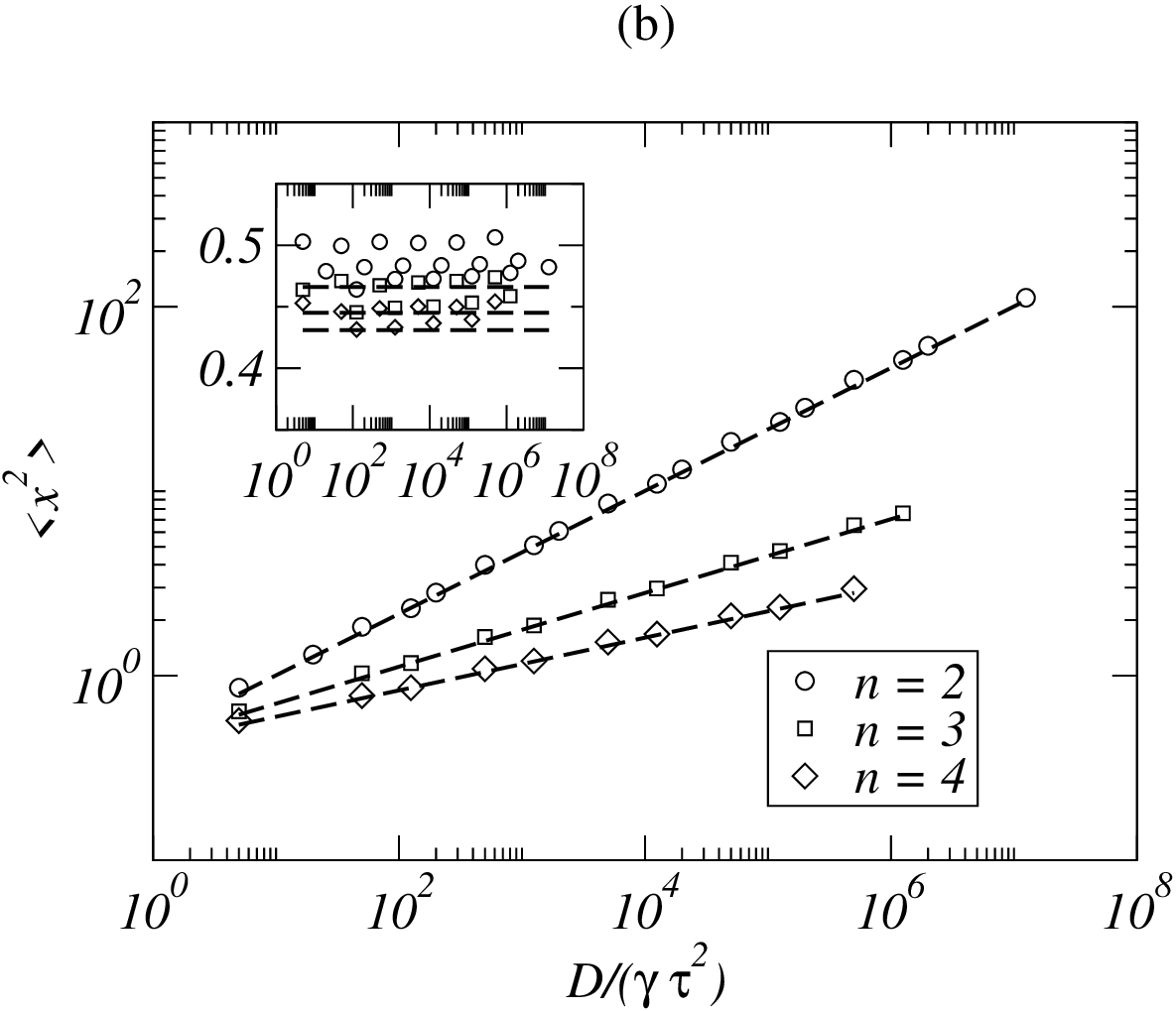}
 \hfill

    \caption{\label{fig:dissip} The 
weakly dissipative case.  
Eq.~(\ref{dissdyn2}) with  Ornstein-Uhlenbeck noise
   is  integrated numerically for $\gamma = 0.02$
with a timestep $\delta t = 10^{-5}$, 
and averaged over $10^3$ realizations. For $n = 2,3,4$, we plot: 
$(a)$ the average $\langle E \rangle$ (main plot) and the ratio 
$\langle E \rangle/(\mathcal{D} /\gamma \tau^2)^{n/(2n-1)}$ (inset)
vs. $\mathcal{D} / \gamma  \tau^2$ ;
$(b)$ $\langle x^2 \rangle$ (main plot) and the ratio 
$\langle x^2 \rangle/(\mathcal{D} /\gamma \tau^2)^{1/(2n-1)}$ (inset)
vs. $\mathcal{D} /\gamma  \tau^2$. The parameters $\tau$ and ${\mathcal D}$
vary within the ranges $2 \le \tau \le 10$, $10^1 \le {\mathcal D} \le 10^6$.
The dashed lines in the figures correspond to the analytical predictions,
Eqs.~(\ref{moystatE}--\ref{moystatx2}). Averages are evaluated beyond 
the dissipative and colored timescales ($t > 1/\gamma$  and 
$t > {\mathcal{D}}^{-1} \; \tau^{-2n/(n-1)}$ respectively). 
}
\end{figure}

 These analytical formulae  compare favorably with numerical
 simulations (Fig.~\ref{fig:dissip}) 
 for parameter values such that both  $\gamma$ and the product
 $\gamma \tau $ are sufficiently small (e.g.,  $\gamma = 0.02$,
 $2 \le \tau \le 10$).
As expected, Eqs.~(\ref{moystatE} to \ref{moystatx2})
 become less and less accurate as the dissipation coefficient 
 $\gamma$ becomes  larger (however,  the anomalous diffusion 
 exponent $n/(2n - 1)$ in (\ref{moystatE}) remains roughly valid 
 at least up to $\gamma \simeq 1$). For smaller values of $\tau$,
 care should be taken to ensure that measurements are indeed
performed beyond the crossover time $t_c$ where white-noise-like
dynamical behavior is replaced by the time-asymptotic regime
characteristic of colored noise 
($t_c \sim {\mathcal{D}}^{-1} \; \tau^{-2n/(n-1)}$ is a 
rapidly decreasing function of $\tau$).

 Lastly, our result for the stationary P.D.F. in the weakly dissipative
 regime,  (Eq.~(\ref{PDFstatE})), compares favorably with the P.D.F. of 
 a linear oscillator with  Ornstein-Uhlenbeck noise \cite{schimansky}
\begin{equation}
    P_{\textrm{stat}}(\dot x, x)  =   
  \frac{ \gamma (1+ \gamma \tau + \tau^2  )  }
   {  \pi {\mathcal D} \sqrt{1+ \gamma \tau }   }
  \exp\Big\{ -\frac{ 2\gamma } {  {\mathcal D}  }
     \Big( \frac{ \dot x^2}{2}  +  \frac{ x^2}{2}
 \frac{ 1}{1+ \gamma \tau }  \Big) (1+ \gamma \tau + \tau^2  ) \Big\}  \, . 
 \label{PDFlinE}
 \end{equation}
 In the limit  $\tau \gg 1 $ and $\gamma \tau \ll 1 ,$
  this expression reduces to 
\begin{equation}
    P_{\textrm{stat}}(E)  = \frac{  2 \gamma \tau^2   } {  {\mathcal D}}
  \exp\Big( -\frac{  2 \gamma \tau^2 E  } {  {\mathcal D}  }
     \Big)  \, , 
 \label{PDFlinE2}
 \end{equation}
 which  agrees with Eq.~(\ref{PDFstatE}) for $n =1.$

  \section{Conclusion}
\label{sec:conclusion}

   We have studied the long  time asymptotic behavior
 of a non-linear oscillator subject to a  Ornstein-Uhlenbeck noise.
 Because of the nonlinearity, the intrinsic  time scale of the system becomes, 
  in the long time limit, 
 smaller than the correlation time of the noise. In this regime,
   the noise is  intrinsically {\it colored} and the behavior
 of the system  is   radically different 
 from that obtained for  the white noise.
 The difference between white and colored
 noise driving can be quantitatively probed by  calculating 
 growth  exponents that characterize the long time behavior 
 of physical observables such as the energy, mean-square amplitude or velocity
 of the oscillator. We   emphasize that the  colored noise scalings
 are  always satisfied in the long time limit however  small  be  the correlation
 time $\tau$  of the noise. However, when $\tau$ is small,  
  a crossover from the white noise scaling regime at short times
 to  colored noise scalings at long times  is observed.  The 
  crossover time $t_c$  diverges to infinity as $\tau \to 0$. In other words, 
  the limits $\tau \to 0$ and $t \to \infty$ ($t$ being the observation
 time) do not commute.

In this work, our analysis
 is based on the exact Fokker-Planck equation obtained  by embedding
 the initial problem  into  a higher dimensional
 Markovian system. In the long time limit, a dynamical separation
  of time-scales occurs: the angular variable becomes a  fast variable  whereas
 the energy evolves on a slower  time scale. It is 
 therefore tempting to eliminate   the   angle and derive an effective
 stochastic dynamics for the slow variable. 
 Straightforward  averaging   leads to 
  excellent results for the white noise case but it fails 
   for  colored noise. Indeed, at the lowest order the
 noise is practically  constant during an angular period: thus, 
   when the  angular degree of freedom is integrated out, 
  the noise itself is  eliminated. 
 A more sophisticated averaging procedure is therefore required
 that embodies the subdominant variations of the noise during
 the short time scale. 
   By making   recursive coordinate  transformations  on this higher
 dimensional system, we have  managed to integrate out  the fast variable
 while retaining  the effect of the noise. We have  finally 
  derived  an effective colored   Langevin dynamics for the 
 slow energy  variable. At leading order, this effective dynamics is 
 solved leading to an  explicit    formula  for the
 probability distribution function  of the energy in the long time limit. 
  From this P.D.F.
  we have  derived  the moments of the position
 and the  velocity of the oscillator.
 Our analytical results  are  compared with  numerical 
 simulations and  the agreement
 is excellent. We may thus  conjecture  (without proof) that
  the analytical expression~(\ref{PDFE}) for  the  P.D.F. of the energy,
  derived  by  our averaging scheme,  is  an  exact  asymptotic result.

  We have extended our analysis to the case 
 where the system is weakly dissipative; again  the agreement between 
 analytical  and  numerical   results  is very good. 
When the dissipation rate becomes higher, the agreement 
 becomes less and less accurate; indeed,  
  the angular variable can no more 
 be considered to be fast   as compared  to the action variable 
 and therefore it is not possible  to use an adiabatic elimination
 procedure.  However,  in the   regime of  very large 
 dissipation rate,  the inertial term  can be adiabatically eliminated  
 \cite{graham,rahman,wu} and the
 problem is reduced to a first order Langevin equation with colored noise; 
 many  specific 
 approximation schemes have been developed
 to study this  case \cite{risken,anishchenko}.  
  The mathematical study of the intermediate dissipation regime
  remains to be done.  Moreover,  the extension of these
 averaging techniques to systems  with spatio-temporal noise
  \cite{vandenb,munoz}  is  a  challenging  open   problem.

\appendix

 \section{Some useful mathematical identities}
\label{sec:app0} 
   In this Appendix, we derive some mathematical relations about
 the hyperelliptic function  ${\mathcal S}_n(\phi)$ that have
 been used in the text.  In Eq.~(\ref{hyperelli}), we defined 
the hyperelliptic function ${\mathcal S}_n$ as
   \begin{equation}
{\mathcal S}_n(\phi) = Y \,  \leftrightarrow 
  \phi     =   \sqrt{n} \int_0^{ \frac{Y}{(2n)^{1/2n}}} 
      \frac{{\textrm d}u}{\sqrt{ 1 - u^{2n}}}  
  =   \frac{ \sqrt{n}} { (2n)^{1/2n} } \int_0^Y 
   \frac{{\textrm d}u}{\sqrt{ 1 -  \frac{u^{2n}}{2n}}}      \,    .
  \label{eq:hyperell}
\end{equation}
      From this equation, we derive the following relation 
 between   ${\mathcal S}_n$ and its derivative ${\mathcal S}_n'$  
\begin{equation}
 {\mathcal S}_n'(\phi) = \frac{ (2n)^{ \frac{1}{2n}}}{\sqrt{n}} \left(
      1 -  \frac{({\mathcal S}_n(\phi))^{2n}}{2n} \right)^{\frac{1}{2}} .
\label{derivS}
\end{equation} 
These hyperelliptic functions reduce to circular and elliptic functions
 for $n=1$  and  $n =2$, respectively: 
\begin{eqnarray}
     {\mathcal S}_1(\phi)   =   \sqrt{2} \sin\phi    \,\,\,\, &\hbox{ and }& \,\,\,\, 
     {\mathcal S}_1'(\phi)  =    \sqrt{2} \cos\phi                        \, ,\\
     {\mathcal S}_2(\phi)   =    \frac{ {\rm sn}\left(\phi;\frac{1}{\sqrt{2}}  \right)}
      { {\rm dn}\left(\phi;\frac{1}{\sqrt{2}}  \right)}   
                     \,\,\,\, &\hbox{ and }&  \,\,\,\,            
    {\mathcal S}_2'(\phi)  =   \frac{ {\rm cn}\left(\phi;\frac{1}{\sqrt{2}} \right)}
      { {\rm dn}^2\left(\phi;\frac{1}{\sqrt{2}}  \right)}                      \, , 
\end{eqnarray}
where sn, cn and dn are the classical Jacobi  elliptic functions
 of modulus $ 1/\sqrt{2}$  \cite{abram,byrd}.

 The functions ${\mathcal S}_n$ and ${\mathcal S}_n'$  are periodic functions
 of period $K_n$, defined in Eq.~(\ref{nperiod}). Their average over
 a period vanishes.  We also 
  need  to calculate  the averages of the square of these functions. 
  Using Eq.~(\ref{eq:hyperell}) 
 and  Eq.~(\ref{nperiod}), we derive Eq.~(\ref{defmu})
 \begin{equation}
   \mu =  \overline{ {\mathcal S}_n^2 } = 
  \frac{1}{K_n}
   \int_0^{K_n} {\mathcal S}_n^2(\phi)\, {\textrm d}\phi = 
  (2n)^{\frac{1}{n}} 
\frac{ \int_0^1 \frac{u^2{\textrm d}u} {\sqrt{ 1 - u^{2n} } } }
    { \int_0^1 \frac{{\textrm d}u} {\sqrt{ 1 - u^{2n} } }  }
 = (2n)^{ \frac{1}{n}} 
    \frac{ \Gamma\Big( \frac{3}{2n} \Big)\Gamma\Big( \frac{n+1}{2n} \Big)     }
       { \Gamma\Big( \frac{1}{2n} \Big)\Gamma\Big( \frac{n+3}{2n} \Big)   }  \, .
\label{eq:moySn}
\end{equation} 
To obtain  this identity,  we made the change of variables  
 $ u = {\mathcal S}_n(\phi) $
  and    then $ r = u^{2n}$, and finally
  used  the Eulerian integral of the first kind \cite{abram}
\begin{equation}
  \int_0^1 r^{\alpha} \left( 1 - r \right)^{\beta} {\textrm d}r
   =   \frac{ \Gamma(\alpha +1)\Gamma(\beta+1) }
       {\Gamma(\alpha + \beta +2) }   \, . 
\end{equation}
Similarly, from Eqs.~(\ref{derivS}) and (\ref{nperiod}),
 and   again making  the change of variable   $ u = {\mathcal S}_n(\phi) ,$
    we deduce that 
\begin{equation}
\overline{ {{\mathcal S}_n'}^2 } 
=  \frac{(2n)^{\frac{1}{n}}}{n } \, \frac{  \int_0^1 \textrm{d}u \, \sqrt{ 1 - u^{2n}}  } 
     {  \int_0^1 \frac{\textrm{d}u}{\sqrt{ 1 - u^{2n}} } }   \, .
 \label{eq:moydSnapp}
 \end{equation}
 Using  the
  following identity (that  can be proved   by integrating
 $ \int_0^1  1. \sqrt{ 1 - u^{2n}} \, \textrm{d}u$ by parts) 
 \begin{equation}
   \int_0^1 \textrm{d}u \, \sqrt{ 1 - u^{2n}} = n \int_0^1
\textrm{d}u \, \frac{u^{2n} }{\sqrt{ 1 - u^{2n}} }  = 
 -n \int_0^1 \textrm{d}u \, \sqrt{ 1 - u^{2n}}
 +n  \int_0^1 \frac{\textrm{d}u}{\sqrt{ 1 - u^{2n}} } \, , 
\label{ipp}
\end{equation} 
we conclude from Eq.~(\ref{eq:moydSnapp}) that 
\begin{equation}
\overline{ {{\mathcal S}_n'}^2 } = \frac{ (2n)^{\frac{1}{n}} }{n+1} \, .
\end{equation}

  \section{Transformation of the original Langevin equations}
\label{sec:app1}

 In this Appendix,  we  perform a change of variables in 
  the original  Langevin equations  and derive 
 the stochastic dynamical  system~(\ref{evZ}, \ref{evphi} and  \ref{evxi}) 
 starting  from  the original equations~(\ref{ev1}, \ref{ev2} and  \ref{ev3}).  
  In terms of the variable $Y$ defined in Eq.~(\ref{defY}), Eq.~(\ref{ev1})
  can be written as  
  \begin{equation}
\frac{\dot  Y}{2n-1}  = \frac{\dot\Omega}{n} 
  \Big ( \frac{\Omega}{ (2n)^{\frac{1}{2n}}} \Big )^{\frac{n-1}{n}} = 
 \Big ( \frac{Y}{ (2n)^{\frac{1}{2n}}} \Big )^{\frac{n-1}{2n -1}}
 {\mathcal S}_n'(\phi)    \, \xi   \, .
 \label{modif1}
  \end{equation}
 Similarly, substituting the variable $Y$ in  Eq.~(\ref{ev2})
  we obtain   
 \begin{equation}
  \dot\phi =
   \Big ( \frac{Y }{ (2n)^{\frac{1}{2n}}} \Big )^{\frac{n-1}{2n - 1}}
   \Big (    1 -  \frac{ {\mathcal S}_n(\phi)  \xi }{ Y }    \Big )  \, . 
 \label{modif2}
  \end{equation}
   From Eqs.~(\ref{modif1} and \ref{modif2}),  we deduce    
 \begin{equation}
\frac{\dot  Y}{2n-1}  = 
 \frac{ {\mathcal S}_n'(\phi)\dot\phi    \, \xi }  
 { 1 -  \frac{ {\mathcal S}_n(\phi)  \xi }{ Y } }     \, .
 \label{modif3}
 \end{equation} 
    At  long times, we know that  $ Y   \to \infty$; thus 
  we  make a perturbative
 expansion of Eq.~(\ref{modif3})  as follows
 \begin{equation}
\frac{\dot  Y}{2n-1}  =  {\mathcal S}_n'(\phi)\dot\phi    \, \xi 
  + \frac{ {\mathcal S}_n'(\phi){\mathcal S}_n(\phi)\dot\phi \, \xi^2 }{ Y }
 + \frac{ {\mathcal S}_n'(\phi){\mathcal S}_n^2(\phi)\dot\phi \, \xi^3 }{ Y^2 }
+ \ldots 
 \label{dvpt1}
\end{equation}
 In the following calculations we  retain   terms up to
 the order  ${\mathcal O}(Y^{-1})$  only,  because $Y \to \infty$
 in mean.  In principle, random fluctuations may allow 
 for small values of $Y$  for arbitrary large times. 
 However, in our numerical calculations, we  never observed  
 such  fluctuations  (of course,  numerical simulations are  
 carried out over a finite duration  of time and  can not be used 
  as a proof that $Y \to \infty$ pathwise). Thus, the effective
 dynamics that we shall derive will be valid   for large values
 of $Y$    and  the P.D.F. that  we shall obtain may not
  describe correctly the distribution of  small values of $Y$.
   Nevertheless, since  the relative probability of  observing
 small values of $Y$ tends to $0$ as time grows,   the 
 formulae for the long time  behavior of the moments   will be 
 asymptotically exact when  $t \to \infty$. 

  Because  we retain only  contributions  
  up to  the order  ${\mathcal O}(Y^{-1})$,  we can neglect    the 
 last term  of Eq.~(\ref{dvpt1}): indeed, 
   from  Eq.~(\ref{modif2})  we observe    that 
 this term  is of  order 
 ${\mathcal O}\Big( \frac{\dot\phi}{ Y^2}  \Big) = 
 {\mathcal O}( Y^{ \frac{1 -3n}{2n -1} }    ) \,  ,$
  which is less than ${\mathcal O}( Y^{-1}  )\, .$
 We now   integrate  by parts   the two relevant
 terms  on   the  right hand side
 of Eq.~(\ref{dvpt1}) in order  to make the  white noise
  contributions appear explicitly:
 \begin{equation}
\frac{\dot  Y}{2n-1}  = 
  \frac{{\textrm d}}{{\textrm d} t}
   \Big(  {\mathcal S}_n(\phi)  \, \xi  +
   \frac{ {\mathcal S}_n^2(\phi) \, \xi^2 }{2  Y }  \Big) 
   +   \frac{{\mathcal S}_n(\phi) (\xi  + \eta)  }{ \tau }
 +  \frac{ {\mathcal S}_n^2(\phi)\, \xi (\xi  + \eta) }{ \tau Y }  
 +   \frac{ {\mathcal S}_n^2(\phi)\, \xi}{ 2 }   \frac{\dot  Y}{Y^2} + \ldots   \, 
 \label{dvpt2}
\end{equation}
  The last term  in  this equation is of the order 
${\mathcal O}\Big( \frac{\dot Y}{ Y^2}  \Big) = 
{\mathcal O}\Big( \frac{\dot\phi}{ Y^2}  \Big)\,  $
 and is negligible with respect to  ${\mathcal O}( Y^{-1}  )\, .$
 We thus obtain up to the desired order 
\begin{equation}
  \frac{{\textrm d}}{{\textrm d} t}  \Big \{   
   \frac{  Y}{2n-1}  -  {\mathcal S}_n(\phi)  \, \xi 
  -  \frac{ {\mathcal S}_n^2(\phi) \, \xi^2 }{2  Y }  \Big \}
  =   \frac{{\mathcal S}_n(\phi) (\xi  + \eta)  }{ \tau }
 +  \frac{ {\mathcal S}_n^2(\phi)\, \xi (\xi  + \eta) }{ \tau Y }   \, .
 \label{dvpt3}
\end{equation}
 On the right hand side  (r.h.s)  of this equation, the  white noise terms 
 (proportional to $\eta$) and 
 the term  ${\mathcal S}_n^2(\phi)\xi^2/ \tau Y$
 will survive the  
  averaging over the  angular variable   $\phi$ whereas 
 the term $ {\mathcal S}_n(\phi)\xi/ \tau $  will be  eliminated
 because $  \overline {{\mathcal S}_n(\phi) } = 0$.
 However, this   term   contains  important  correlations
 between    $\xi$  and  $\phi$ of the  order  ${\mathcal O}(Y^{-1})$
 and in order to retain 
  these correlations,  we transform  this   term   as follows 
 \begin{eqnarray}
  \frac{{\mathcal S}_n(\phi)\xi}{\tau} &=& \frac{{\mathcal S}_n(\phi)\xi}{\tau}
  \frac{\dot\phi}{\dot\phi} =
  \frac{  {\mathcal S}_n(\phi) \dot\phi \, \xi  } 
 {  \tau \Big ( \frac{Y }{ (2n)^{\frac{1}{2n}}} \Big )^{\frac{n-1}{2n - 1}}
   \Big (    1 -  \frac{ {\mathcal S}_n(\phi)  \xi }{ Y }    \Big ) } 
   \nonumber  \\ 
   &=&
\frac{ (2n)^{\frac{n-1}{2n(2n - 1)}}  {\mathcal S}_n(\phi) \dot\phi  \, \xi   } 
{\tau \,  Y^{\frac{n-1}{2n - 1}}} 
 + 
\frac{  (2n)^{ \frac{n-1}{2n(2n - 1)}    }  {\mathcal S}_n^2(\phi) \dot\phi  \, \xi^2   } 
{\tau \,   Y^{\frac{3n-2}{2n - 1}}} + \ldots  \, 
  \label{dvpt4}
\end{eqnarray}
  Integrating   by parts the first term on the r.h.s.  of
 Eq.~(\ref{dvpt4}) and using   the function  ${\mathcal C}_n(\phi)$
   defined in Eq.~(\ref{defCn}), we    obtain
 \begin{eqnarray}
 \frac{  
  (2n)^{ \frac{n-1}{2n(2n - 1)} }  
  {\mathcal S}_n(\phi) \dot\phi  \, \xi   } 
{\tau \,  Y^{\frac{n-1}{2n - 1}}}  &=& 
 \frac{{\textrm d}}{{\textrm d} t}   \Big ( 
  \frac{  (2n)^{ \frac{n-1}{2n(2n - 1)} }    {\mathcal C}_n(\phi)   \, \xi   } 
{\tau \,  Y^{\frac{n-1}{2n - 1}}}\Big )  +
  \frac{  (2n)^{ \frac{n-1}{2n(2n - 1)} }      {\mathcal C}_n(\phi)   \, ( \xi + \eta)   } 
{ \tau^2  \,  Y^{\frac{n-1}{2n - 1}}}
   +  
 \frac{(2n)^{ \frac{n-1}{2n(2n - 1)} }  (n-1) {\mathcal C}_n(\phi) \, \xi \, \dot Y   } 
{  (2n - 1) \tau \,    Y^{\frac{3n-2}{2n - 1}}}    \nonumber  \\
 &=& \frac{{\textrm d}}{{\textrm d} t}   \Big ( 
  \frac{  (2n)^{ \frac{n-1}{2n(2n - 1)} }     {\mathcal C}_n(\phi)   \, \xi   } 
{ \tau \, Y^{\frac{n-1}{2n - 1}}}\Big )  +
  \frac{  (2n)^{ \frac{n-1}{2n(2n - 1)} }  {\mathcal C}_n(\phi)   \, ( \xi + \eta)   } 
{ \tau^2 \,   Y^{\frac{n-1}{2n - 1}}}
   +  \frac{ (n-1)  \, {\mathcal S}_n'(\phi) {\mathcal C}_n(\phi)   \, \xi^2   } 
{ \tau \,   Y} \,  ,  \label{trf1}
   \end{eqnarray} 
where we have used  Eq.~(\ref{modif1}) to derive the last equality. 
Similarly, the second term  on the r.h.s.  of
 Eq.~(\ref{dvpt4}) is rewritten as
\begin{equation}
\frac{  (2n)^{ \frac{n-1}{2n(2n - 1)}}  {\mathcal S}_n^2(\phi) \dot\phi  \, \xi^2   } 
{\tau \,   Y^{\frac{3n-2}{2n - 1}}}
  = 
\frac{ {\mathcal S}_n^2(\phi)  \, \xi^2   } {\tau \,  Y }
  \frac{  \dot\phi  } {\Big ( \frac{Y }{ (2n)^{\frac{1}{2n}}} \Big )^{\frac{n-1}{2n - 1}}}
  = \frac{ {\mathcal S}_n^2(\phi)  \, \xi^2 } {\tau \, Y } + {\mathcal O}( Y^{-2} ) \, , 
  \label{trf2}
\end{equation}
where we have used  Eq.~(\ref{modif2}). If we substitute 
 Eqs.~(\ref{trf1} and \ref{trf2}) in  Eq.~(\ref{dvpt4})
 and  then substitute the result in Eq.~(\ref{dvpt3}), we derive the following
 equation for $Y$, valid up to  the order  
 ${\mathcal O}( Y^{-1}  )$
\begin{eqnarray}
  &&\frac{{\textrm d}}{{\textrm d} t}  \Big \{   
   \frac{ Y}{2n-1}  -  {\mathcal S}_n(\phi)  \, \xi 
 - \frac {  (2n)^{\frac{n-1}{2n(2n-1)}}   {\mathcal C}_n(\phi) \xi }
  { \tau  Y^{\frac{n-1}{2n -1}}  } 
  -  \frac{ {\mathcal S}_n^2(\phi) \, \xi^2 }{2  Y }  \Big \}
  =   \nonumber 
 \\ &&  \frac {  (2n)^{\frac{n-1}{2n(2n-1)}}   {\mathcal C}_n(\phi) \xi }
  { \tau^2  Y^{\frac{n-1}{2n -1}}  } 
 + \frac{   2 {\mathcal S}_n^2(\phi)\, +  (n-1)\, {\mathcal S}_n'(\phi) {\mathcal C}_n(\phi)   } 
{ \tau  Y}  \, \xi^2   \nonumber  \\ &&   + \frac{1}{ \tau }  \eta(t)
  \Big \{   {\mathcal S}_n(\phi) +  
\frac {  (2n)^{\frac{n-1}{2n(2n-1)}}   {\mathcal C}_n(\phi) \xi }
  { \tau  Y^{\frac{n-1}{2n -1}}  } 
 +  \frac{ {\mathcal S}_n^2(\phi)\, \xi  }{  Y }   \Big \} \, .
 \label{EqY}
\end{eqnarray}
 The left hand side of  Eq.~(\ref{EqY}) is identical to 
 $\frac{ \dot Z}{2n-1}$, with 
   $Z$   defined in  Eq.~(\ref{defZ}).
 On the  right  hand side of  Eq.~(\ref{EqY}), we  must  
  express  $Y$  as a function of   $Z$. Although 
 the relation~(\ref{defZ}) between $Z$ and $Y$  can not be inverted  by a closed formula,
  $Y$ can be calculated  as a function of $Z$ up to  the 
  order ${\mathcal O}( Z^{-1})$ included.   This  leads 
 to Eq.~(\ref{evZ}) and to  the formulae~(\ref{courZ} and \ref{diffZ})  
 for the current and diffusion functions. 

  \section{Derivation of the averaged Fokker-Planck equation}
 \label{sec:app2}

 In this Appendix, we  derive the averaged Fokker-Planck equation~(\ref{FPZav})
 by eliminating the  fast  angular variable from   Eq.~(\ref{FPZ}).
 We  assume  that the P.D.F.   $\Pi_t(Z,\phi,\xi)$  becomes
 independent  of   $\phi$ when $ t \to \infty$,
 {\it i.e.},   the   probability measure becomes
 uniform with respect to the angular  variable 
in the long time limit. We   integrate  Eq.~(\ref{FPZ})
  with respect to  the fast  angular variable $\phi$ and examine the behavior
  of each term of  Eq.~(\ref{FPZ}). 
 The average of the current
$J_Z(Z,\phi,\xi)$ is given by:
\begin{equation}
  \overline{J_Z} =  \frac {  (2n)^{\frac{n-1}{2n(2n-1)}} 
 \overline{{\mathcal C}_n(\phi) }\, \xi }
  { \tau^2  Z^{\frac{n-1}{2n -1}}  }  + 
  \frac{ 2 \overline{{\mathcal S}_n^2(\phi) } + 
 (n-1) \overline{ {\mathcal S}_n'(\phi) {\mathcal C}_n(\phi)} }{ \tau Z} \xi^2 \, .
\end{equation}
Integrating by parts  and using   Eq.~(\ref{defCn}), we obtain  
 $$ \overline{ {\mathcal S}_n'(\phi) {\mathcal C}_n(\phi)}  =
   - \overline{ {\mathcal S}_n(\phi) {\mathcal C}_n'(\phi)} 
 = -  \overline{{\mathcal S}_n^2(\phi) } \, .  $$
Using  Eq.~(\ref{defmu}),  we deduce  that 
\begin{equation}
  \overline{J_Z} = \frac { 3 - n } { \tau Z} \, \mu  \xi^2 \, .
\label{JZav}
\end{equation}
 Therefore, we have 
\begin{equation}
 \overline{ \frac{\partial }{\partial Z }  (J_Z \Pi_t)  }   =  
 \frac{\partial }{\partial Z }     \Bigg( \frac{ (3 -n)\xi^2  } 
 {\tau Z}  \tilde\Pi_t  \Bigg)   \, . 
\label{dJZdZav}
\end{equation}
Similarly,  the average of the diffusion term 
 ${\mathcal D}_Z(Z,\phi,\xi)$ is given by:
 \begin{equation}
\overline{ {\mathcal D}_Z}= \overline{ {\mathcal S}_n(\phi) } 
 +  \frac {  (2n)^{\frac{n-1}{2n(2n-1)}}   \overline{  {\mathcal C}_n(\phi) }   }
  { \tau  Z^{\frac{n-1}{2n -1}}  }   +
\frac { \overline{  {\mathcal S}_n^2(\phi) } \, \xi  }{Z} 
 =  \frac {  \mu  \xi }{Z}  \, , 
\label{DZav}
\end{equation}
where we have used  
 $ \overline{ {\mathcal S}_n } = \overline{ {\mathcal C}_n } = 0 .$
   The angular average of the expressions
$\frac{\partial^2 }{\partial Z\partial \xi   }
  ({\mathcal D}_Z  \Pi_t)$
 and     $\frac{\partial }{\partial Z}
 {\mathcal D}_Z \frac{ \partial  \Pi_t}{\partial \xi}$ 
 that appear in  Eq.~(\ref{FPZ}) are readily deduced by using  Eq.~(\ref{DZav}):  
\begin{eqnarray}
  \overline{ \frac{\partial^2 }{\partial Z\partial \xi }  ({\mathcal D}_Z  \Pi_t)} 
  &=&      \mu \frac{\partial^2  }
 {\partial Z\partial \xi   }   \Bigg(\frac{  \xi \tilde\Pi_t }{Z} \Bigg)    \, ,
 \label{dZdxiav1}  \\
\overline{   \frac{\partial }{\partial Z}
 {\mathcal D}_Z \frac{ \partial  \Pi_t}{\partial \xi} }  &=&   
  \mu \frac{\partial }{\partial Z} \Bigg(   \frac{\xi}{Z} 
  \frac{ \partial \tilde\Pi_t}{\partial \xi} \Bigg)    \, . 
\label{dZdxiav2}
 \end{eqnarray}
 
  Recalling that $\frac{\partial }{\partial Z}$ scales as $Z^{-1}$,
  the leading order in the  expression
  $\frac{\partial }{\partial Z} {\mathcal D}_Z \frac{\partial }{\partial Z}
  ({\mathcal D}_Z  \Pi_t)$  scales as   $Z^{-2}$. Therefore,  
   retaining   only  the leading term in the  average of  this  expression,  
  we obtain 
 \begin{equation}
\overline{ \frac{\partial }{\partial Z} {\mathcal D}_Z \frac{\partial }{\partial Z}
  ({\mathcal D}_Z  \Pi_t)}  = \mu \frac{\partial^2 \tilde\Pi_t  }{\partial Z^2} \, .
 \label{d2Z2av}
\end{equation}

 Finally, using  the fact that  the integral  of any  term of
 the type $\partial_\phi(\ldots)$  over a period of $\phi$   vanishes,
 we deduce
\begin{equation}
\overline{ \frac{\partial }{\partial \phi} (J_\phi \Pi_t) } = 0 \, .
 \label{dJphi}
\end{equation}

 We have thus calculated the  angular average of each term
 that appears on the right hand side of Eq.~(\ref{FPZ})
  [see Eqs.~(\ref{dJZdZav}, \ref{dZdxiav1}, \ref{dZdxiav2}, \ref{d2Z2av} 
  and \ref{dJphi})]. 
 This concludes the
 derivation of Eq.~(\ref{FPZav}).

 \section{The `Best  Fokker-Planck Equation' approximation}
 \label{sec:appBFPE}

   In this appendix, we shall discuss  an approximation called the 
   `Best  Fokker-Planck Equation'  (B.F.P.E.)   which is based
   on a partial resummation of the  perturbative small $\tau$ 
   expansion  \cite{lindcol1,lindcol2}. 
  This B.F.P.E. has been rightly criticized
  \cite{marchesoni}   because it can 
  lead to quantitatively incorrect results ({\it e.g.}, in  the
  study of the activation rate of  overdamped bistable oscillators
  driven by colored noise). Nevertheless, we shall
  show that,  for the system under study, 
  the B.F.P.E. leads to  the   correct colored 
  noise scaling behavior; it can also  describe  the  crossover 
  between white and colored noise regimes as we showed in \cite{philkir4}.
  Because the B.F.P.E. can not
  be expected to yield   exact  quantitative results, our discussion 
  will be   mostly qualitative.

    Because random  dynamical systems  with colored noise are  non-Markovian, 
    there exists no  closed Fokker-Planck equation 
   that describes the dynamics of the P.D.F. in  phase-space.
   The Markovian character can be retrieved by embedding the system
   into a higher dimensional process, as 
   has been  done in  the present work (see Section \ref{sec:averaging}).
   Nevertheless,  many approximation schemes have been developed 
   to derive effective  Fokker-Planck equations for systems subject
   to colored noise   such as short correlation time expansions
     \cite{sancho1,sancho2,weinstein},  functional methods \cite{fox,tsironis},
   unified colored noise approximation (U.C.N.A.)\cite{schimansky},
 or  decoupling Ansatz \cite{moss}
 (for a review, see \cite{hanggirev1,hanggirev2,hanggirev3}).
     We  have compared  some  of these  approximations   with
   our averaging technique and our numerical results. 
   We  have  observed that no  short correlation time expansion 
   truncated at any finite order could  yield  the correct colored
   noise scalings of Eqs.~(\ref{scalingcolor}). Because, 
  in the long time limit,  the correlation
  time $\tau$ of the noise is not the shortest time scale of  the system,
   any finite perturbative expansion with respect to $\tau$ is bound to fail.
  An  expansion that embodies terms of all orders is needed:  an  approximate 
     resummation of this type  is provided by the  B.F.P.E.

     The  B.F.P.E. is constructed from an  
     expansion of the stochastic Liouville equation associated
     with the system~(\ref{evolomega} and \ref{evolphi}).  
    These   equations  are  rather  involved  due to the presence of
    hyperelliptic functions and  because of the  indirect coupling
    between $\Omega$ and $\phi$ through the noise term $\xi$.
    We can  radically    simplify  these equations without
     altering their qualitative properties  by writing 
   \begin{eqnarray}
     \dot \Omega  &=&  \kappa  \cos \phi   \, \xi(t)  
 \label{appomega}
   \,  ,   \\   
 \dot\phi  &=& \Big ( \frac{\Omega}{ (2n)^{\frac{1}{2n}}} 
\Big )^{\frac{n-1}{n}}   \, .
    \label{appphi}
   \end{eqnarray}
   We have neglected the $1/\Omega$ term with respect to
   $\Omega^{(n-1)/n}$  in the equation for $\phi$. Moreover, 
    the hyperelliptic function   ${\mathcal S}_n'(\phi)$   has been replaced 
    by  the   ordinary  circular function   $\kappa  \cos \phi$.
   The constant  $\kappa$  will be adjusted later so that
 the white noise limit of the system~(\ref{appomega} and \ref{appphi})
  coincides with    that of the original 
   system~(\ref{evolomega} and \ref{evolphi}).

   The second-order cumulant expansion  \cite{vankampen} 
  of the stochastic Liouville equation corresponding
  to Eqs.~(\ref{evolomega} and \ref{evolphi}) is an approximate
  evolution equation for the P.D.F.  $P_t$  in the small noise limit.
  It is given by 
 \begin{equation}
\frac{ \partial   P_t }{\partial t}  = 
 {\bf L}_0 P_t  + \int_0^t dx \langle {\bf L}_1(t) \exp({\bf L}_0x)
    {\bf L}_1(t-x)   \exp(-{\bf L}_0x) \rangle P_t  \, ,  
\label{FPcumul}
\end{equation}
where  the differential operators are defined as 
 \begin{eqnarray}
    {\bf L}_0  P_t &=&  - \frac{\partial}{\partial \phi}
  \Big(  \big ( \frac{\Omega}{ (2n)^{\frac{1}{2n}}} 
\big )^{\frac{n-1}{n}}  P_t  \Big)  \, ,
\label{defL0}  \\
    {\bf L}_1(t) P_t &=&  - \frac{ \partial}{\partial \Omega}
 ( \xi(t)  \kappa  \cos \phi\,  P_t) \,.\label{defL1}
  \end{eqnarray}
 Following the procedure  of  \cite{lindcol1}, we 
  evaluate the right hand side of Eq.~(\ref{FPcumul}) by applying 
  the  following  operator formula
 \begin{equation}
 \exp(A) B \exp(-A) =  B + [A,B] + \frac{1}{2!} [A, [A,B]] + 
 \frac{1}{3!} [A, [A, [A,B]]] + \ldots  \,, 
\label{idmat}
\end{equation}
 with  $ A = {\bf L}_0$   and $B = {\bf L}_1(t-x)$; we must therefore
 calculate iterated commutators of the operators ${\bf L}_0$  
 and  ${\bf L}_1$. By induction, it can be shown that 
 the $n$-th  commutator  is of the form 
\begin{equation}
 T_k = 
[{\bf L}_0,[\ldots,[{\bf L}_0, [{\bf L}_0, {\bf L}_1(t-x)]]\ldots]  =   \kappa  \xi(t-x)
  \left( \frac{ \partial   }{\partial \Omega}  H_1^{(k)}(\Omega,\phi)
   +  \frac{ \partial   }{\partial \phi}  H_2^{(k)}(\Omega,\phi)    \right)
 \, , \label{defTn}
\end{equation}
where the  functions $H_1^{(k)}$   and $H_2^{(k)}$
  satisfy the  recursion relations   
\begin{eqnarray}
   H_1^{(k)} &=& -  \Big( \frac{\Omega}{ (2n)^{\frac{1}{2n}}} \Big)^{\frac{n-1}{n}} 
   \frac{ \partial H_1^{(k-1)}}{\partial \phi} \, ,  \\
   H_2^{(k)} &=& {\frac{n-1}{n}}
    \frac{   \Omega^{-\frac{1}{n}}   }{ (2n)^{\frac{n-1}{2n^2}}} H_1^{(k-1)} 
  -  \Big( \frac{\Omega}{ (2n)^{\frac{1}{2n}}} \Big)^{\frac{n-1}{n}} 
   \frac{ \partial H_2^{(k-1)}   }{\partial \phi} \, ,
\label{recurrence}
\end{eqnarray}
 and the initial conditions $H_1^{(0)} = -\cos \phi $, $H_2^{(0)} = 0$.
 The solution  of these recursions is readily  guessed once the first
 few terms have been calculated and we obtain 
\begin{eqnarray}
   H_1^{(k)} &=&  (-1)^{k-1}
 \Big( \frac{\Omega}{ (2n)^{\frac{1}{2n}}} \Big)^{\frac{(n-1)k}{n}} 
   \cos(\phi + k \frac{\pi}{2})    \, , \label{solrecurrence1} \\
   H_2^{(k)} &=&    (-1)^{k} \frac{(n-1)k}{n \Omega  } 
 \Big( \frac{\Omega}{ (2n)^{\frac{1}{2n}}} \Big)^{\frac{(n-1)k}{n}} 
  \sin(\phi +  k \frac{\pi}{2})  
  \, . 
\label{solrecurrence2}
\end{eqnarray}
  Using  the expressions~(\ref{solrecurrence1})~and~(\ref{solrecurrence2})
  for  $H_1^{(k)}$ and $ H_2^{(k)}$, respectively,   and  the autocorrelation
 function~(\ref{deftau}) of the Ornstein-Uhlenbeck noise, 
 we calculate the integral on  the right hand side of Eq.~(\ref{FPcumul}) 
\begin{eqnarray}
&& \int_0^t {\rm d}x \langle {\bf L}_1(t) \exp({\bf L}_0x) 
     {\bf L}_1(t-x)   \exp(-{\bf L}_0x) \rangle    \nonumber \\
&&  =  -\frac{\mathcal D \kappa  \cos\phi  }{2 \, \tau}   \,
 \left(  \frac{\partial^2}{\partial \Omega^2} {\mathcal H}_1(\Omega,\phi,t)
     +  \frac{ \partial^2   }{\partial \Omega\partial \phi}
    {\mathcal H}_2(\Omega,\phi,t)  \right)  \, ,
\label{devrhs}
 \end{eqnarray}  
where we have defined 
\begin{equation} 
{\mathcal H}_1(\Omega,\phi,t)   = 
  \sum_{k=0}^{\infty} \frac{ \int_0^t {\rm d}x  x^k {\rm e}^{-x/\tau } }{k!}
  H_1^{(k)}   \,\,\,\,  \hbox{ and } \,\,\,\,   
 {\mathcal H}_2(\Omega,\phi,t)   = 
  \sum_{k=0}^{\infty} \frac{ \int_0^t {\rm d}x  x^k {\rm e}^{-x/\tau } }{k!}
  H_2^{(k)}   
 \label{defgotH2}
\end{equation}
 In the limit $t \to \infty$, the integral  
$\int_0^{{t}/{\tau}} {\rm d}x \,  x^k {\rm e}^{-x }$ converges
 to $ k!$,   and the series defining  
${\mathcal H}_1$ and ${\mathcal H}_2$  can be calculated 
in a closed form~: 
\begin{eqnarray} 
{\mathcal H}_1(\Omega,\phi,\infty)  &=& 
  - \tau 
 \frac{ \cos\phi +  \tau \omega  \; \sin\phi}
{ 1 + \tau^2 \omega^2 }  \, ,\\
  {\mathcal H}_2(\Omega,\phi,\infty)  &=& 
  - \tau^2   \frac{n-1}{n}\;
 \frac{  \Omega^{-\frac{1}{n}}  }{ (2n)^{\frac{n-1}{2n^2}}} \;
\frac{ (1 -  \tau^2 \omega^2  )
 \cos\phi  + 2  \tau \omega \; \sin\phi } 
 { (1 + \tau^2 \omega^2 )^2}
 \, .
\end{eqnarray}
where we have defined 
\begin{equation}
  \omega =   \Big( \frac{\Omega}{ (2n)^{\frac{1}{2n}}} \Big)^{\frac{n-1}{n}}\, .
\end{equation}
Substituting these expressions in Eq.~(\ref{devrhs}),  we derive  the
 B.F.P.E. associated with the system~(\ref{appomega} and \ref{appphi}) 
\begin{eqnarray}
 \frac{ \partial   P_t }{\partial t}  =  
 &-&  \frac{\partial}{\partial \phi}
  \Big(  \big ( \frac{\Omega}{ (2n)^{\frac{1}{2n}}} 
\big )^{\frac{n-1}{n}}  P_t  \Big) 
  +  \frac{ {\mathcal D} \kappa }{2 } 
   \frac{\partial^2}{\partial \Omega^2} 
\left(  \frac{ \cos^2\phi  + \tau \omega  \; \sin\phi\cos\phi}
 { 1 + \tau^2\omega ^2} P_t  \right)   \nonumber \\ 
  &+& \frac{ {\mathcal D} \kappa \tau \cos\phi }{2 }  \frac{n-1}{n}
 \frac{\partial^2}{\partial \Omega\partial \phi}
  \left(   \frac{  \Omega^{-\frac{1}{n}}  }{ (2n)^{\frac{n-1}{2n^2}}} \;
\frac{ (1 -  \tau^2 \omega^2  )
 \cos\phi  + 2  \tau \omega \; \sin\phi } 
 { (1 + \tau^2 \omega^2 )^2}
 P_t  \right) \, .  \nonumber \\  
   \label{bfpe}
\end{eqnarray}
  Integrating   out   the fast variable $\phi$, we obtain 
 an averaged B.F.P.E.  for ${\tilde P}_t(\Omega)$,
the  probability  distribution of 
 the slow variable, 
  \begin{equation}
  \frac{ \partial   {\tilde P}_t }{\partial t}  = 
  \frac{ {\mathcal D \kappa  }  }{4} 
   \frac{\partial}{\partial \Omega}  \Big(
   \frac{1} {  1 + \tau^2 \omega^2}
  \frac{\partial {\tilde P}_t }{\partial \Omega}   
  \Big) \, .
   \label{avbfpe}
\end{equation}
 For $\tau =0$,  this equation leads to a scaling
 behavior identical  to   the averaged white
 noise Fokker-Planck equation (\ref{nFPmoy})
 and the average of $\Omega^2$ agrees with 
 Eq.~(\ref{scalingomegablanc})  if we choose
 \begin{equation}
  \kappa =  (2n)^{\frac{n+1}{n}} \, .
 \end{equation}
For a non-zero $\tau$, the scaling behavior predicted
 by the averaged  B.F.P.E. (\ref{avbfpe}) is indeed
 $\Omega \sim  t^{n/(4n -2)} $ which is in  agreement
  with Eq.~(\ref{scalingomegacolor}). For   $\tau \ll  1$,
  Eq.~(\ref{avbfpe}) describes the crossover between the
  white noise regime at short times and 
 the colored noise regime at long times \cite{philkir4}.

\end{document}